\documentclass[11pt,a4paper,english,pdftex]{article}
\usepackage{a4wide}
\usepackage{cite} 
\usepackage{url}  
\usepackage{ifthen}  
\usepackage{multicol}   
\usepackage{bm, color, epsfig,graphicx,subfigure,graphics,amsmath,amssymb,amsthm,amsfonts, colordvi, url,rotating}
\usepackage[utf8]{inputenc} 
\usepackage{algorithm,algorithmic}
\usepackage{rotating}

\newcommand{\cor}{\operatorname{cor}}

\newcommand{\sX}{{\sf{X}}}
\newcommand{\sZ}{{\sf{Z}}}

\newcommand{\bbeta}{\mbox{\boldmath$\beta$}}

\newcommand{\bSigma}{\mbox{\boldmath$\Sigma$}}

\newcommand{\bOmega}{\mbox{\boldmath$\Omega$}}

\bmdefine\c{c} \bmdefine\C{C} \bmdefine\U{U} \bmdefine\X{X}
\bmdefine\D{D} \bmdefine\K{K} \bmdefine\Z{Z} \bmdefine\x{x}
\bmdefine\z{z} \bmdefine\y{y} \bmdefine\Y{Y}
\bmdefine\bfalpha{\alpha} \bmdefine\bfmu{\mu} \bmdefine\M{M}
\bmdefine\Q{Q} \bmdefine\P{P} \bmdefine\w{w} \bmdefine\W{W}
\bmdefine\p{p} \bmdefine\T{T} \bmdefine\t{t} \bmdefine\r{r}
\bmdefine\a{a} \bmdefine\B{B} \bmdefine\I{I} \bmdefine\u{u}
\bmdefine\p{p} \bmdefine\Sig{\Sigma} \bmdefine\E{E} \bmdefine\F{F}
\bmdefine\S{S} \bmdefine\s{s} \bmdefine\w{w} \bmdefine\b{b}
\bmdefine\W{W} \bmdefine\w{w} \bmdefine\V{V} \bmdefine\v{v}
\bmdefine\q{q} \bmdefine\R{R} \bmdefine\A{A} \bmdefine\H{H}
\bmdefine\d{d} \bmdefine\g{g} \bmdefine\L{L}

\newtheorem{axiom}{Axiom}

\newtheorem{definition}[axiom]{Definition}

\theoremstyle{definition}

\begin{document}

\title{Regularized estimation of large-scale gene association networks using graphical Gaussian models}

\author{Nicole Kr\"{a}mer\\Machine Learning Group\\ Berlin Institute of Technology\\\texttt{nkraemer@cs.tu-berlin.de}  \and Juliane Sch\"{a}fer\\Basel Inst. for Clinical Epidemiology \& Biostatistics\\University Hospital Basel\\\texttt{jschaefer@uhbs.ch}  \and Anne-Laure Boulesteix\\Department of Medical Informatics, Biometry \& Epidemiology\\University of Munich\\\texttt{boulesteix@ibe.med.uni-muenchen.de}}


\maketitle

\begin{abstract}
Graphical Gaussian models are popular tools  for the estimation of (undirected) gene association networks from microarray data.
A key issue when the number of variables greatly exceeds the number of samples is the estimation of the matrix of partial correlations.
Since the (Moore-Penrose) inverse of the sample covariance matrix leads to poor estimates in this scenario, standard methods are inappropriate and adequate regularization techniques are needed.
Popular approaches include biased estimates of the covariance matrix and high-dimensional regression schemes, such as the Lasso and Partial Least Squares.\\
In this article, we investigate a general framework for combining regularized regression methods with the estimation of Graphical Gaussian models. This framework includes various existing methods  as well as two new approaches based on ridge regression and adaptive lasso, respectively. These methods are extensively compared both qualitatively and quantitatively within a simulation study and through an application to six diverse real data
sets. In addition, all proposed algorithms are implemented in the R package ``parcor'', available from the R repository CRAN.
In our simulation studies, the investigated non-sparse regression methods, i.e. Ridge Regression and Partial Least Squares, exhibit rather conservative behavior when combined with (local) false discovery rate multiple testing in order to decide whether or not an edge is present in the network. For networks with higher densities, the difference in performance of the methods decreases.
For sparse networks, we confirm the Lasso's well known tendency towards selecting too many edges, whereas the two-stage adaptive Lasso is an interesting alternative that provides sparser solutions. In our simulations, both sparse and non-sparse methods are able to reconstruct networks with cluster structures.
On six real data sets, we also clearly distinguish the results obtained using the non-sparse methods and those obtained using the sparse methods where specification of the regularization parameter automatically means model selection.
Furthermore, for data that violate the assumption of uncorrelated observations (due to replications), the Lasso and the adaptive Lasso yield very complex structures, indicating that they might not be suited under these conditions. The shrinkage approach is more stable than the regression based approaches when using subsampling.
\end{abstract}
\section{Introduction}
Besides Bayesian networks \cite{Friedman0401}, auto-regressive models \cite{Yeung0201}, and state-space models \cite{Rangel0401}, graphical Gaussian models (GGMs) are a popular method for modeling genetic networks based on microarray transcriptome data.
In the GGM methodology \cite{Whittaker9001}, which is considered in the present article, networks are represented as undirected graphs. Each vertex represents a gene, and an edge connects two genes if they are partially correlated. In contrast to correlation, which measures both direct and indirect interactions between pairs of variables, partial correlation measures the strength of direct interaction only.
Since investigators are primarily interested in direct gene interactions, the GGM framework is attractive for modeling of regulatory networks: several recent methodological articles report successful applications
of GGMs to the estimation of genetic networks from microarray data \cite{Dobra0401,Schaefer0501,Schaefer0502,Li0601,Yuan0701,Pihur0801,Friedman0801}.
These approaches are used in numerous applied studies, e.g., for estimating {\it Arabidopsis} gene networks \cite{Ma0701} or for the study of genetically mediated cortical networks \cite{Schmitt0801}.

Nonetheless, reconstructing GGMs from high-dimensional microarray data remains a difficult task. The standard estimation of partial correlations involves either the inversion of the sample covariance matrix, or the estimation of $p$ least squares regression problems, where $p$ is the number of genes. If the number $n$ of observations (arrays) is much smaller than the number $p$ of variables (genes), these approaches are inappropriate. Suitable alternatives are based either on regularized estimation of the (inverse) covariance matrix, or on regularized high-dimensional regression. The present paper focuses on the latter approach, and presents a comparative study on the use of various approaches to high-dimensional regression for covariance selection. The chosen methods are extensively compared in simulations and real data studies. Since for real data the ground truth (i.e. the true underlying network) is unknown, our performance analysis focuses on the similarities and differences between the investigated methods. In particular, we examine the connectivity and size of the resulting graphs, as the differences between the estimated networks. Moreover, we compare the stability of the methods with respect to  subsampling and with respect to violations of i.i.d. assumptions.

In the remainder of this section, we give a brief overview of graphical Gaussian modeling in the classical setting with $n > p$. Subsequently, we discuss the case of high-dimensional data in the ''Methods'' section.

\subsection{Gene Regulatory Networks and Graphical Gaussian Models}
\label{subsec:grn}

Graphical Gaussian models (GGMs) \cite{Whittaker9001} are
fundamental tools in order to represent direct covariate interactions.
Formally, a GGM is an undirected graph whose nodes represent variables, and whose edges represent conditional dependency relations. An edge between two nodes is missing if and only if they are conditionally independent given all other nodes.  Assuming a joint normal distribution, the conditional dependence can be quantified in terms of partial correlations.
For a random variable $\sX$ and a finite set of
random variables $\mathcal{Z}=\{{\sZ_1},\ldots,{\sZ_k}\}$,
the orthogonal complement of ${\sX}$  with respect to
$\mathcal{Z}$ is
\begin{eqnarray*}
{\sX}_{\setminus \mathcal{Z}} &=& {\sX} - \mathcal{P}_{\mathcal{Z}} {\sX},
\end{eqnarray*}
where the projection $\mathcal{P}_{\mathcal{Z}}$ is defined with respect to the inner product
$\langle {\sX_1}, {\sX_2} \rangle = E[{\sX_1} {\sX_2}]$ between two random variables $\sX_1$ and $\sX_2$.  Here, we tacitly assume that all involved moments exist. The partial correlation $\rho_{\mathcal{Z}}\left({\sX_1},{\sX_2}\right)$ between ${\sX_1}$ and ${\sX_2}$ with respect to
$\mathcal{Z}$ is the correlation of the orthogonal complements of ${\sX_1}$ and
${\sX_2}$ with respect to $\mathcal{Z}$:
\begin{eqnarray}
\label{eq:par}
\rho_{\mathcal{Z}}\left({\sX_1},{\sX_2}\right) &=&
\cor \left( {\sX_1}_{\setminus \mathcal{Z}},{\sX_2}_{\setminus \mathcal{Z}} \right).
\end{eqnarray}
In the context of gene regulatory networks, each of the $p$ genes is represented by a random variable $\sX_i$ ($i=1,\dots,p$). For each pair  of genes $(i,j)$, we are interested in their partial correlation $\rho_{ij}$  with respect to all other genes, i.e. with respect to the set of random variables $\mathcal{Z}_{\setminus ij} =
\left\{{\sX_1},\ldots,{\sX_p}\right\} \setminus \{{\sX_i},{\sX_j}\}$.

Given $n$ observations (arrays) $\x_1,\ldots,\x_n \in \mathbb{R}^p$ of the set of $p$ genes, the standard unbiased plug-in estimate for the partial correlation coefficients $\rho_{ij}$ in the case $n>p$ can be formulated in two equivalent ways \cite{Whittaker9001}, as outlined below.

\subsubsection*{Notations}

In the rest of this article,
\begin{eqnarray}
\label{eq:X}
\X &=& \left(\x_1, \ldots,\x_n\right)^\top \in \mathbb{R}^{n \times p}
\end{eqnarray}

denotes the $n\times p$ column-centered data matrix with rows corresponding to observations (arrays) and columns corresponding to variables (genes). The standard unbiased estimate of the $p \times p$ covariance matrix $\bSigma$ is then given as
\[
\widehat \bSigma =\frac{1}{n-1} \X^\top \X.
\]

\subsubsection*{Formulation 1: Inversion of the Covariance Matrix}

If the estimate $\widehat \bSigma$ is invertible,  an unbiased estimate of the  partial correlation between genes $i$ and $j$ is obtained as
\begin{eqnarray}
\label{eq:rho_inv} \widehat \rho_{ij}&=& -\frac{\widehat
\omega_{ij}}{\sqrt{\widehat \omega_{ii}\, \widehat \omega_{jj}}}\,.
\end{eqnarray}
with $\widehat \bOmega$ denoting the inverse of the estimated covariance matrix:
\[
\widehat \bOmega= (\widehat{\omega}_{ij})=\widehat \bSigma
^{-1}.
\]

\subsubsection*{Formulation 2: Least Squares Regression}

Let us consider the $p$ linear regression models
\begin{eqnarray}
\label{eq:regmodel} {\sX_i} &=&  \sum_{j\not =
i} \beta^{(i)} _j {\sX_j} + \varepsilon\,,\ \mbox{for\ }i=1,\dots,p,
\end{eqnarray}
where $\varepsilon$ stands for i.i.d. noise. Note that we do not include an intercept in the model because the variables are centered. For $i=1,\dots,p$, the least squares estimate $\widehat \bbeta ^{(i)}=(\beta^{(i)} _1,\ldots,\beta^{(i)}_{i-1},\beta^{(i)}_{i+1},\dots,\beta^{(i)}_{p})^\top$ of the vector of regression coefficients is the solution of the optimization problem
\begin{eqnarray}
\label{eq:ols1}
\widehat \bbeta ^{(i)}&=& \text{arg}\min_{\bbeta \in \mathbb{R}^{p-1}} \left \|\X^{(i)} - \X^{(\setminus i)}\bbeta\right \| ^2\\
\label{eq:ols2} &=& \left( \X^{(\setminus i)\top} \X^{(\setminus i)} \right)^{-1} \X^{(\setminus i) \top} \X^{(i)},
\end{eqnarray}
where $\X^{(i)}\in\mathbb{R}^n$ is the $i$th column of $\X$ and $\X^{(\setminus i)}\in\mathbb{R}^{n\times (p-1)}$ is the matrix obtained from $\X$ by deleting the $i$th column. The partial correlation  between genes $i$ and $j$ is then estimated as
\begin{eqnarray}
\label{eq:rho_reg} \widehat \rho_{ij}&=& \text{sign}\left( \widehat
\beta^{(i)} _j \right) \sqrt{\widehat \beta^{(i)} _j \widehat
\beta^{(j)} _i}\,.
\end{eqnarray}
In the $n>p$ setting, the two regression coefficients $\widehat{\beta}^{(i)} _j$ and $\widehat{\beta}^{(j)} _i$ always have the same sign. Hence, $\sqrt{\widehat \beta^{(i)} _j \widehat
\beta^{(j)} _i}$ is well-defined.
Moreover, it can be shown that both formulations 1 and 2 are equivalent \cite{Whittaker9001} in the sense that they always yield the same estimate. In the $n \ge p$ setting, a test of the null hypothesis $\rho_{ij}=0$ is available using results on the distribution of $\widehat \rho_{ij}$.

In microarray data, the number $n$ of samples
is typically very small as compared to the number $p$ of considered genes. Hence, the above framework is inappropriate for two reasons.
Firstly, the standard estimate of the partial correlation matrix given by Eqs. (\ref{eq:rho_inv}) and (\ref{eq:rho_reg}) is not appropriate when $n<p$: in formulation $1$, the estimated covariance matrix $\widehat\bSigma$ is typically ill-conditioned or even singular, and its generalized (Moore-Penrose) inverse has large mean squared error \cite{Schaefer0501}. In formulation $2$, the least squares criterion (\ref{eq:ols1}) is ill-posed and leads to overfitting. Hence, an alternative regularized estimate of the partial correlation matrix has to be used in the context of GGMs with high-dimensional data. The two formulations 1 and 2 lead to two different strategies for the regularized estimation of the partial correlations in the $p \gg n$ setting, which are reviewed in the Methods section.

Secondly, the testing approach mentioned above breaks down in the $p \gg n$ setting, since the sampling distribution of estimates $\widehat{\rho}_{ij}$ under the null hypothesis of zero partial correlation is unknown. Two alternatives have been proposed in order to assess statistical significance: (i) methods  based on sparse estimates of the partial correlation matrix that do not require separate testing, and (ii) methods based on empirical null modeling and (local) false discovery rate multiple testing \cite{Efron0403,Schaefer0502,Strimmer0802}.

\section{Methods}

This section reviews the available strategies for estimating GGMs in the $p \gg n$ setting: biased large-scale covariance estimation and  regularized regression including our two novel variants (Ridge Regression and Adaptive Lasso).

\subsection{Regularized Estimation of the (Inverse) Covariance Matrix}
\label{subsec:regcorr}

This approach is derived from formulation 1. The general approach is to plug a regularized estimate of the inverse
of the sample covariance matrix  into Eq. (\ref{eq:rho_inv}).
Sch\"afer \& Strimmer \cite{Schaefer0501} adopt this approach and propose a  shrinkage estimator of the covariance matrix.  This shrinkage estimator is constructed as a convex combination of the unrestricted sample covariance matrix $\widehat\bSigma$ and an estimator $\widehat \T$ of a specified low-dimensional sub-model $\T$:
\begin{eqnarray*}
\widehat\bSigma_{\lambda} = \lambda \widehat \T + (1- \lambda) \widehat\bSigma,
\end{eqnarray*}
where the factor $\lambda \in [0,1]$ controls the shrinkage intensity. Let us assume a parametrization of covariances in terms of correlations and variances, whereas shrinkage is applied to the correlations and diagonal entries are left intact, i.e. the estimator does not shrink the variances.  For correlation shrinkage, we consider the identity matrix as the most commonly employed shrinkage target. Notice that the optimal shrinkage intensity $\lambda$ can be determined analytically and be estimated from the data. Thus, the resulting correlation shrinkage estimator is positive definite, and favorable properties carry over to derived quantities, such as sample partial correlations. Subsequently, model selection of the gene association network can be achieved using empirical null modeling and (local) false discovery rate multiple testing \cite{Efron0403,Schaefer0502,Strimmer0802}.

Estimates of the inverse covariance matrix can also be obtained using bootstrap aggregating (bagging) as a technique for variance reduction \cite{Breiman9601}. In some implicit way, the bootstrap procedure presumably helps to regularize the problem. However, bagging schemes are inferior to the shrinkage estimator \cite{Schaefer0501}, and computationally much more expensive.  A recent extension using the augmented bootstrap \cite{Tyekucheva0801} is in fact closely related to the shrinkage estimator \cite{Strimmer0801,Schaefer0801} and is expected to perform similarly.

In this paper, we use the correlation shrinkage based approach as a reference method in comparison with the regression based approaches to covariance selection.

Finally, recent novel approaches are to be noted that are based on $\ell_1$ regularized maximum likelihood estimation in graphical Gaussian models \cite{Yuan0701,dAspremont0801,Friedman0801,Rothman-etal08,witten09}.  Corresponding inverse covariance estimates exploit the sparsity in the graphical structure and conduct parameter estimation and model selection simultaneously.  However, despite recent advances in semidefinite programming computation remains challenging in practice due to the high-dimensionality and positive definiteness constraint \cite{Yuan08}.

\subsection{Regularized Regression}
\label{subsec:regreg}

Here, the strategy is to replace the least squares estimator in (\ref{eq:ols2}) by some regularized estimator of the regression coefficients that can be used in formula (\ref{eq:rho_reg}) to obtain estimators of the partial correlations. More formally, we define the following class of estimates of the partial correlations.

\begin{definition}
\label{def:parcor}

For any regression method \texttt{reg} that yields (regularized) estimates $\widehat \bbeta ^{(i)} _{\operatorname{reg}}$ of the linear regression model (\ref{eq:regmodel}), we define the corresponding estimate of the partial correlations as
\begin{eqnarray*}
\widehat \rho_{ij,\operatorname{reg}}&=&
\operatorname{sign}\left(\widehat \beta^{(i)} _{j,\operatorname{reg}}\right)  \min \left \{1, \sqrt{\widehat
\beta^{(i)} _{j,\operatorname{reg}} \widehat \beta^{(j)} _{i,\operatorname{reg}} }\right\}
\end{eqnarray*}
\begin{eqnarray*}
\mbox{if}\ \operatorname{sign} \left(\widehat \beta^{(i)} _{j,\operatorname{reg}}\right)&=& \operatorname{sign}\left(\widehat \beta^{(j)} _{i,\operatorname{reg}}\right)
\end{eqnarray*}
and $0$ otherwise.
\end{definition}

This definition ensures that the estimated partial correlation coefficients are always well-defined and that they lie in the interval $[-1,1]$. Again, we can roughly distinguish between regression methods that require testing to construct the undirected graphs, and sparse regression methods.

In the rest of this subsection, we discuss two regularized regression methods (PLS and the Lasso) that have been proposed for the estimation of large-scale GGMs in the literature.  Furthermore, we propose two additional attractive methods (ridge regression and the adaptive Lasso).

\subsubsection*{Partial Least Squares}

Tenenhaus et.~al. \cite{Tenenhaus0801} suggest Partial Least Squares (PLS) regression \cite{Wold7501,Wold8401} as a plug-in for Def. \ref{def:parcor}. PLS is a method for supervised dimensionality reduction. It has its seed in the chemometrics community, but its success has lead to applications in various other scientific fields, e.g. in chemo- and bioinformatics \cite{Saigo0801,Boulesteix0701}. The main idea of PLS  is to build a few
orthogonal components  from the original data $\X^{(\setminus i)}$ and to use them as
predictors in a least squares fit. A PLS component  $\t = \X^{(\setminus i)} \w  $ is a linear
combination of the original predictors that  have maximal covariance with the
 response vector $\X^{(i)}$, under the additional assumption that the components are mutually orthogonal. Formally, the $k$-th PLS component is defined by
\begin{eqnarray*}
\w_k &=& \text{arg}\max_{\|\w\|=1}\text{cov}\left( \X^{(\setminus i)} \w,\X^{(i)} \right)^2\\
\text{s.t. } &&\w ^\top \X^{(\setminus i) \top } \X^{(\setminus i)} \w_l =0 \,\, \text{ for }l<k\,.
\end{eqnarray*}

Hence, PLS regularizes the regression problem by compressing the $p$ variables into a small number $m$ of orthogonal components $\T=\left(\t_1,\ldots,\t_m\right)$ and regressing the response variable onto these components. After rescaling the weight vectors $\w_k$ ($k=1,\dots,m$) such that $\t_k$ has length 1, this leads to the regression coefficients
  \begin{eqnarray*}
  \widehat \bbeta ^{(i)}_{\text{pls}}&=& \left(\w_1,\ldots,\w_m\right) \T^\top \X^{(i)}\,.
  \end{eqnarray*}

While the original formulation of PLS  scales with the number $p$ of variables, it is also possible  to represent the algorithm in a way that it only scales with the number $n$ of observations \cite{Rosipal0101,Rosipal0601}. This leads to a  dramatic decrease in computation time for $p \gg n$.  Note that the number of PLS components is a model parameter that has to be optimized for each of the $p$ regression models (\ref{eq:regmodel}).  The standard model selection techniques are cross-validation or information criteria based on degrees of freedom \cite{Kraemer0701}. In the context of gene regulatory networks, Tenenhaus et.al. \cite{Tenenhaus0801} propose to use the same number of components $m$ for all $p$ regression models. They observe empirically that  the partial correlation coefficients (Def. \ref{def:parcor}) obtained from PLS regression reach a plateau when the number of PLS components $m$ increases, and suggest a heuristic procedure to choose the smallest $m$ for
 which the plateau is reached. However, in our experiments, we use the theoretically well-funded and popular cross-validation technique with $k$ folds.

As the PLS coefficients are not sparse, the obtained partial correlations are in general non-zero. Thus, a statistical testing procedure has to be used to determine which edges are significant. (Alternatively, one might also use a sparsification of PLS as proposed by Chun \& Keles \cite{chun09}.) In the present article, we use large-scale simultaneous hypothesis testing with local false discovery rate (fdr) level 0.2, in order to identify unusual outliers among the estimated partial correlations.

For the sake of completeness, let us mention in this section a variant of the PLS approach described above, which was recently suggested by Pihur et al. \cite{Pihur0801}. Instead of estimating the partial correlation using Eq. (\ref{eq:rho_reg}), they propose an alternative measure of correlation strength which is very similar to the PLS-based partial correlation coefficient except that, roughly speaking, the square root of the product of $\widehat{\beta}_{j,pls}^{(i)}$ and $\widehat{\beta}_{i,pls}^{(j)}$ is replaced by their sum. We remark that Pihur et.~al. do not optimize the number of PLS components $m$ and recommend to use $m \approx 3$.

\subsubsection*{Ridge Regression}

Ridge regression (see e.g. \cite{Hoerl0001}) is probably the most popular and most straightforward regularized regression technique.
Regularization is performed by adding a penalty term $P(\bbeta)$ to the least squares criterion (\ref{eq:ols1}). Ridge regression is based on an $\ell_2$ penalty term of the form

\begin{eqnarray}
\label{eq:ridge}
P(\bbeta) = \lambda \left \| \bbeta\right \|_2^2=\lambda \sum_{i}\beta_i^2,
\end{eqnarray}

where $\lambda>0$ denotes the penalty parameter.  This leads to a reduction of variance and thus avoids overfitting.

The solution obtained by ridge regression depends on the penalty parameter $\lambda$.  In our paper, we use standard $k$-fold cross-validation to select the optimal amount of penalization $\lambda$.
As ridge regression does not lead to sparse solutions, we use large-scale false discovery rate multiple testing \cite{Strimmer0802} to test for significant edges, as described above in the subsection on PLS. Again, we adopt a level of $0.2$.

\subsubsection*{The Lasso}

Meinshausen and B\"{u}hlmann \cite{Meinshausen0601} propose to estimate the regression coefficients in Def. \ref{def:parcor} with the Lasso \cite{Tibshirani9601} and study under which conditions model selection consistency applies, hinging on the choice of the penalty.  Similarly to Ridge Regression, the estimated regression coefficients are chosen to minimize a  penalized least squares criterion.  Lasso regression is based on a $\ell_1$-penalty of the form
\begin{eqnarray}
\label{eq:lasso}
P(\bbeta) = \lambda \left \| \bbeta\right \|_1= \lambda \sum_{i} \left |\beta_i\right|,
\end{eqnarray}
where $\lambda>0$ is the regularization parameter. With the $\ell_1$-penalty, many estimated regression coefficients will be equal to $0$. As a result, with variable selection in mind, the Lasso has a major advantage: a sparse estimator of the matrix of partial correlations is yielded and a graph can be obtained by assigning an edge between two genes if and only if $\widehat \rho_{ij,\operatorname{lasso}}\neq 0$.
The choice of the penalty $\lambda$ has to be determined for each of the $p$ high-dimensional regressions successively.  Again, this can be done using some cross-validation scheme or information criteria.  Meinshausen \& B\"{u}hlmann \cite{Meinshausen0601} motivate a choice of the penalty parameter that aims at controlling the probability of falsely connecting two nodes in the graph, i.e. that is a choice tailored to the graph structure. However, experiments \cite{Schaefer0501} indicate that this approach leads to graphs that are too dense, i.e. too many edges are selected. Therefore, in this paper, we use the oracle penalty for optimal prediction that is determined using $k$-fold cross-validation.

\subsubsection*{The two-stage adaptive Lasso}

The Lasso is only asymptotically consistent for covariance selection when requiring certain necessary conditions among the variables in the GGM. Zhou et al. \cite{Zhou0901} show that the two-stage adaptive Lasso procedure \cite{Zou0601} is consistent for high-dimensional model selection in graphical Gaussian models under rather general and less restrictive conditions.
The adaptive Lasso \cite{Zou0601} considers the Lasso with penalty weights as

\begin{eqnarray}
\label{eq:weighted.lasso}
P(\bbeta) = \lambda \sum_{i} \widehat w_i  \left|\beta_i  \right|,
\end{eqnarray}

where the weights $\widehat w_i$ are chosen in a data-dependent manner.  Specifically, the adaptive Lasso is defined as follows.  Suppose $\widehat \bbeta$ is a $\sqrt{n}$ consistent initial estimator of $\bbeta$.  For example, we can use the least squares estimator $\widehat \bbeta_{\operatorname{ols}}$.  Pick a $\gamma > 0$, and define the weights $\widehat w_i = 1/|\widehat \beta_{i, \operatorname{ols}}|^\gamma$.  The most common choice is $\gamma = 1$.   Here, we use the Lasso estimator $\widehat \bbeta_{\operatorname{lasso}}$ as initial estimator, and define the weights

\begin{eqnarray*}
\widehat w_i  = 1/|\widehat \beta_{i, \operatorname{lasso}}|.
\end{eqnarray*}

Note that the amount of penalization in both the initial stage Lasso and the second stage Lasso with penalty weights is determined via $k$-fold cross-validation.  The adaptive Lasso will be at least as sparse as the Lasso.  For graphical Gaussian modeling, the adaptive Lasso estimates are used in Def. \ref{def:parcor}, and two genes are connected if and only if the  partial correlation coefficient $\widehat \rho_{ij,\operatorname{adaptive\ lasso}} \ne 0$. We remark that for model selection, the optimal weights have to be determined in each of the $k$ cross-validation splits. As the optimal weights themselves are determined via $k$-fold cross-validation, this implies that a lasso fit has to be computed $k^2$ times! This leads to high computational costs.

\section{Results}
\label{sec:results}

In this section, we perform extensive experiments to compare regression-based methods for reconstructing gene regulatory networks.  We consider the recently proposed techniques PLS regression and Lasso regression, and the two additional methods, ridge regression and adaptive Lasso regression, that have not been applied in practice for this purpose before.
As a reference method, we use the  shrinkage approach to covariance estimation, followed by matrix inversion.  An overview of the five considered methods and their respective parameters and characteristic features is given in Table 1.  All methods are implemented in the R package ``parcor'' \cite{parcor}, available from the R repository CRAN.

\subsection{Simulations}
\label{subsec:simulations}

The performance of the  proposed methods is assessed in a simulation study with a set-up similar to \cite{Schaefer0501}.  The number of variables is fixed at $p=100$, and various sample sizes ranging from $25$ to $200$ in steps of $25$ are investigated. We consider two different scenarios. First, we simulate networks with varying degree of density and no network topology, and second, we investigate sparse networks with different network topologies. These scenarios correspond to particular choices of the partial correlation matrix $\P$ (see below). For all experiments, a total of $20$ replications are performed for each sample size to average out variability due to random sampling.  For each replication, the data are drawn randomly from a multivariate normal distribution with correlation structure derived from $\P$.
\subsubsection*{Varying degree of density}
Partial correlation matrices $\P$ of size $p\times p$ with a proportion of
\begin{eqnarray*}
d&=&5\%;10\%;15\%;20\%;25\%
\end{eqnarray*}
non-zero entries are constructed by first drawing the non-zero entries from a uniform distribution on $[-1,1]$ and then rescaling the non-diagonal entries to ensure that we obtain a feasible partial correlation matrix (for more details, see the \texttt{R}-package \texttt{GeneNet} \cite{Schaefer0801}). Hence, the range of the non-zero partial correlations depend on the density of the network. If the network is rather dense, the absolute values of the non-zero partial correlation coefficients are very small compared to a sparse network. This is illustrated in the supplementary file \texttt{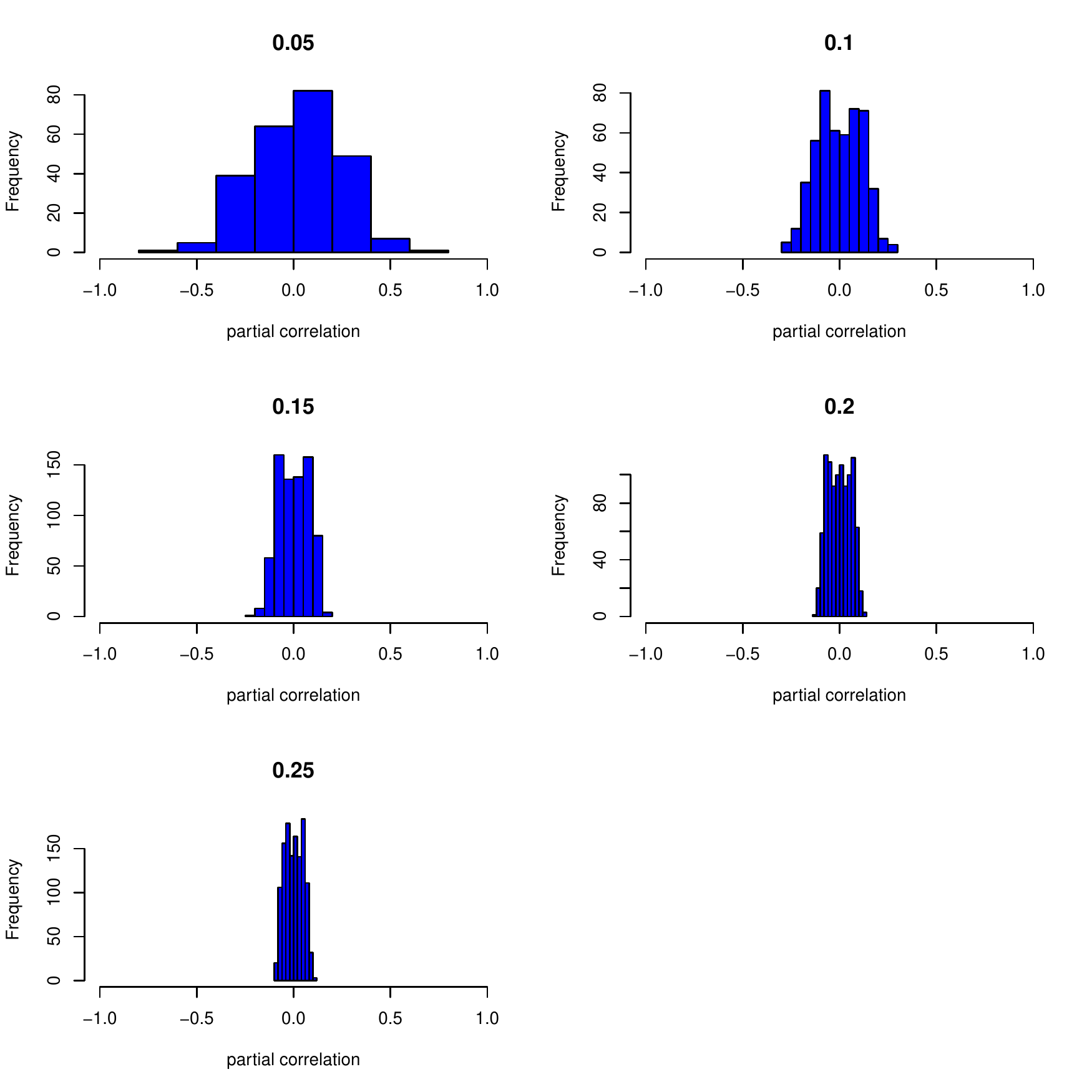}. Here, we plot the histogram of the non-zero partial correlations for a random matrix $\P$ of density $d$. It is important to note that due to the small values, the reconstruction of the network becomes more delicate for a higher degree of density: it is more difficult to select the correct non-zero entries if their true vales are close to zero. We remark that this effect cannot be entirely eliminated by a more clever simulation design, and that the simulation of partial correlation matrices is far from trivial \cite{Ruschhaupt08}.

For each generated data set, $\P$ is then estimated based on PLS regression, ridge regression, the Lasso, the adaptive Lasso and the shrinkage covariance estimator, successively.  For all regression-based methods, $k=5$-fold cross-validation is used to optimize the model parameters, i.e. the number of components $m$ for PLS and the penalty $\lambda$ for ridge regression, the Lasso and the two-stage adaptive Lasso, respectively.  For the Lasso and the adaptive Lasso, we follow the parametrization implemented in the \texttt{lars} package \cite{lars}, based on the ratio of the $\ell_1$-norm of the Lasso and the $\ell_1$-norm of the least squares estimates.  Specifically, the regularization parameter is chosen from an equidistant sequence between $0$ and $1$ of length $1000$. Furthermore, we normalize this parameter to avoid the peaking phenomenon at $n=p$ (see \cite{Kraemer0901} for details).  For ridge regression, we consider a logarithmically spaced sequence $l_1,\dots,l_{1000}$   ranging from $10^{-10}$ to $10^{-1}$.  The candidate penalty parameters are then defined as $\lambda_s=l_s \,n \,p$ (with $s=1,\dots,1000$).  Finally, the range of the number of PLS components is from $1$ to $15$.

We evaluate the accuracy of the resulting estimators  in two respects:  (i) the estimation error of the partial correlation matrix itself, and (ii) the recovery of the underlying networked topology.  The difference between the estimated and true matrix of partial correlations is measured in terms of the mean squared error (MSE).  In the upper left panel of Figures 1a-e, the MSE is displayed as a function of the sample size $n$.

For sparse networks, the two sparse estimates based on the Lasso and the adaptive Lasso, respectively, yield a lower MSE compared to the three other methods that are not sparse and are likely to contain many non-zero but non-significant (small) entries, which ultimately lead to a higher MSE. This effect vanishes for higher degrees of density. A notable exception is PLS. For denser networks, the MSE becomes larger. These networks correspond to small absolute values of the entries in $\P$. Therefore, we conjecture that PLS is not able to shrink the regression coefficients enough, as the regularization parameter $m$ (number of components) is discrete. This is in contrast to the four other methods. Note however that  for the reconstruction of the underlying networked topology the MSE is only of secondary interest.

For each investigated sample size, the resulting number of selected edges is displayed in the upper right panel of Figures 1a-e, while the horizontal line
is the number of true edges.  For sparse networks, the Lasso with its regularization parameter chosen to be prediction optimal tends to select too many edges.  PLS, ridge regression and the approach based on shrinkage covariance estimation are in contrast far more conservative and rather select too few edges, even in the $n>p$ case.  The adaptive Lasso is less conservative and appears to be a promising alternative. Again, these differences vanish for higher degrees of densities. As remarked above, the reconstruction task becomes more difficult for higher degrees of density. This explains the low number of selected edges for higher degrees of density.

The two lower panels in Figures 1a-e correspond to the power (left) and the true discovery rate (tdr, right) which are defined as

\begin{eqnarray*}
\operatorname{power} & = & \frac{\#\{\text{true edges that are selected}\}}{\#\{\text{true edges}\}}\quad \operatorname{and}\\
\operatorname{tdr}   & = & \frac{\#\{\text{true edges that are selected}\}}{\#\{\text{selected edges}\}},
\end{eqnarray*}

respectively.  The panels illustrate that for sparse networks, the Lasso's comparatively high power comes at the prize of rather low true discovery rate. Again, the power decreases with the increase in density of the network. In many practical applications, we argue that it might be more valuable to report more stable results with fewer false positives.

However, it is to be noted that the non-sparse methods using fdr-based procedures for edge selection involve an arbitrary parameter: the fdr threshold (here 0.2). These methods can thus be made more or less sparse by changing the threshold value. To investigate the relative accuracy of the non-sparse methods independently of the particular fdr threshold, the same simulations are subsequently performed with other thresholds. In order to evaluate the ability of the three methods to detect non-zero partial correlations, their sensitivity and specificity are computed for these different fdr thresholds and displayed graphically in form of ROC curves. These can be found in the supplementary material. PLS and ridge regression yield very similar results. They slightly outperform the approach based on shrinkage covariance estimation.  The sensitivity and specificity of the Lasso and the adaptive Lasso, which do not depend on a particular threshold, are depicted as single points.  They are above the ROC curves of the three non-sparse methods, indicating good performance -- especially for the adaptive Lasso.

Finally, we compare the runtime of the respective methods in Figure 2. Note that we do not display the runtime of the Lasso, as it was computed as an intermediate step in the \texttt{R}-function for the adaptive Lasso. The upper part of Figure 2 clearly shows that the computational load for the adaptive Lasso is very high. This is due to the fact that we have to run the lasso algorithm $k^2$ times in $k$-fold cross-validation, and that the (adaptive) lasso algorithm scales unfavorably in the number of variables -- in contrast to PLS, Ridge Regression or shrinkage. The lower part of Figure 2 only displays the runtime of the three latter methods. Shrinkage is faster than the regression based approaches as it circumvents both time-consuming cross-validation and the computation of $p$ different regression models. The discrepancy with respect to the runtime becomes even more apparent in the real data study (see below).

\subsubsection*{Different network topologies}
Next, we consider different network topologies. We simulate two different types of topologies, which are displayed in the supplementary file \texttt{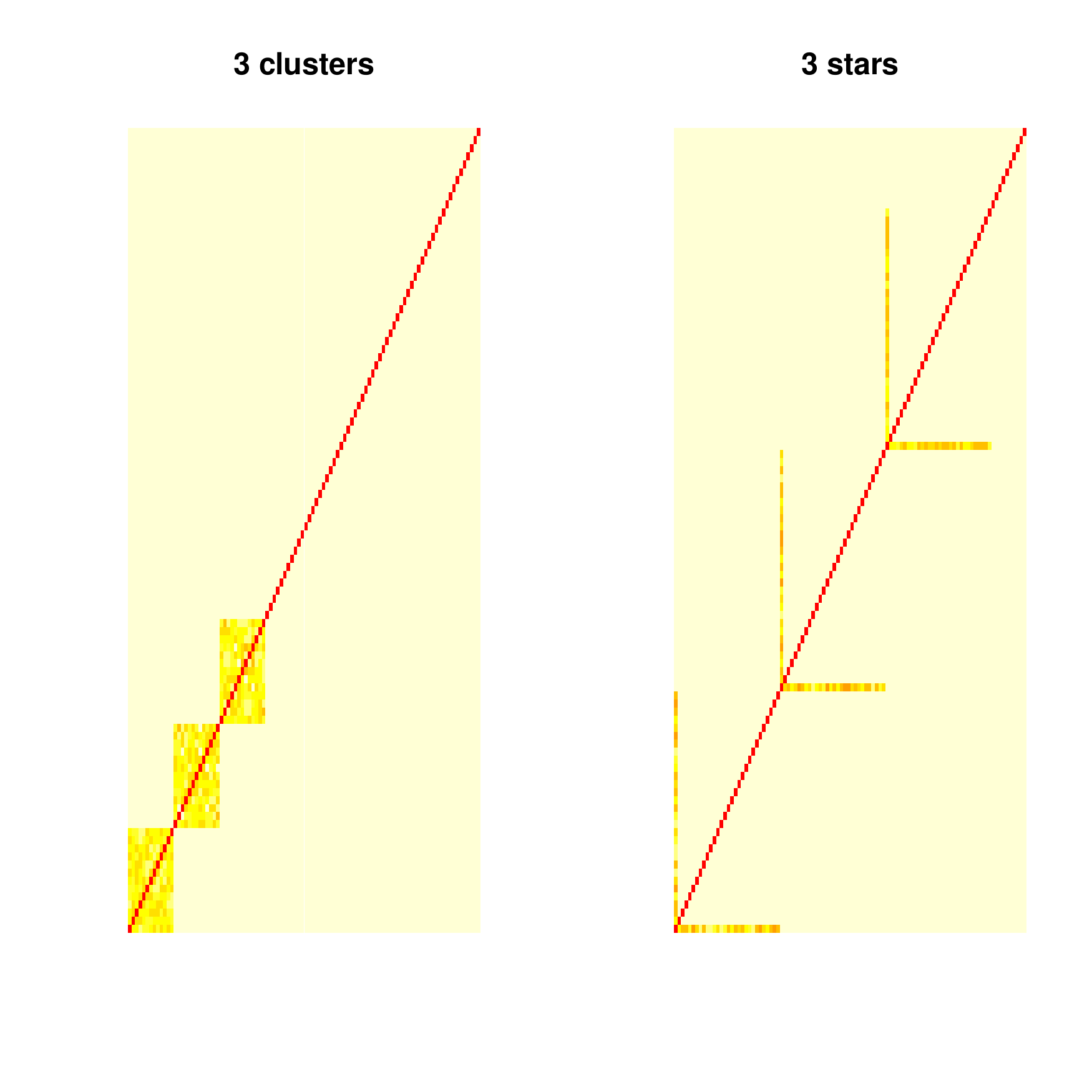}. The left part of the figure  shows three clusters of genes. In each cluster, all genes are partially correlated, and genes from different clusters are not partially correlated. In the simulation, we consider networks with 1,2 and 3 clusters. The right part of the figure in \texttt{clusters.pdf}  shows three star-shaped clusters. In each star, all genes are partially correlated to one gene, the center of the star. In the simulation, we consider a network with 3 stars. The MSE, the number of selected edges, the power and the true discovery rate are displayed in Figures 3a-d. Again, we observe a high MSE for PLS in most scenarios. As explained above, this is probably due to the insufficient shrinkage of PLS towards 0. Overall, the Lasso and Ridge Regression perform best in these scenarios. So, in contrast to what is often conjectured/reported in the literature, we do find in our simulations that sparse methods are able to reconstruct networks in the presence of cluster structures.

\subsection{Real Data Study}
\label{subsec:realdata}

We compare the five different methods on diverse real world data sets: the \texttt{ecoli1} \cite{Kao0401} and \texttt{ecoli2} \cite{schmidt04}, \texttt{Ara} \cite{Smith0401}, \texttt{t.cell10} and \texttt{t.cell34} \cite{Rangel0401}, and \texttt{west} \cite{West0101} data sets. All data sets are freely available. An overview of the size, characteristics and availability of the data sets is given in Table 2.
The five considered methods (shrinkage covariance estimation, ridge regression, PLS, Lasso, adaptive Lasso) including the model selection procedures for the regression-based approaches  are exactly as in the simulation setting.  For \texttt{ecoli2}, we use leave-one-out-cross-validation for model selection, and for \texttt{west}, we use $k=5$-fold cross-validation. For the remaining $4$ data sets, we use $k=10$.

In real world scenarios, the ground truth, i.e. the true underlying network, is hardly ever known, and the performance of different methods cannot be determined in terms of MSE, power and tdr as in the simulation study. Nevertheless, it is possible to compare the performance of the different methods quantitatively. In particular, we investigate the size and the connectivity of the estimated graphs, their overlap the type of interaction between genes and their stability.

Figures 4a and 4b display the percentage of selected edges for each data set. As in the simulation study, the proportion of selected edges strongly depends on the chosen estimation method. More surprisingly, the relative levels of sparsity of the obtained graphs show very different patterns for the six investigated data sets.
The Lasso and adaptive Lasso seem to behave very differently from the other methods. This can at least partly be explained by the fact that they rely on a completely different edge selection scheme which essentially depends on the sparsity of the regression method and not on the testing scheme.

In a nutshell, the Lasso and adaptive Lasso select less edges than the other methods for all data sets except for the two data sets \texttt{t.cell10} and \texttt{t.cell34} with repeated measurements.  With these two data sets, Lasso and adaptive Lasso yield complex graphs with as much as over 50 \% non-zero edges (\texttt{t.cell34} data).  This behavior is likely to be due to the longitudinal structure of the data that is not explicitly considered, since the standard Lasso regression method assumes independent observations.  In contrast, longitudinal structures may be handled in some implicit way by methods using an fdr-based assessment, where the distribution under the null hypothesis is estimated from the data.

PLS reconstructs a very dense network for the data sets \texttt{ecoli1}, \texttt{ara} and \texttt{west}. In combination with the high MSE that we observed in the simulations, we conjecture that PLS in combination with cross-validation is not the most reliable method for the reconstruction of networks. We believe that other model selection strategies or the incorporation of sparse PLS \cite{chun09} may improve the performance of PLS.

Among the three methods with fdr-based assessment of the edges, i.e PLS, ridge regression and shrinkage covariance estimation, the latter procedure seems to be most conservative, whereas PLS identifies the highest number of edges.  This result is consistent for all six real data sets and yields a refinement of the results presented in the simulation study, where these three methods performed similarly.


Table 3 displays the overlap of the estimated graphs. The estimated graphs show a moderate  overlap between the methods. While considering these results, one should keep in mind that
the proportions of selected edges vary a lot across the five methods, which of course decreases the overlap considerably: a very sparse graph can obviously include only a very small proportion of the edges of a more complex graph. Interestingly, the overlap seems to be higher on average for the \texttt{west} data set including the highest number of genes than for the other five data sets.
We remark that the Lasso and adaptive Lasso solutions are computed based on different, random cross-validation splits. This explains that, in  general, the graph found by adaptive Lasso is \emph{not} exactly a subgraph of the solution found by Lasso.

Figures 5a and 5b display the connectivity of the estimated graphs for each of the six data sets. For each gene, we derive the proportion of genes that are connected to it through an edge, with each of the six data sets and each of the five methods. Each boxplot depicts the distribution of the proportion of connected genes for the considered method and the considered data set.
As explained above, the assumption of i.i.d. observations is violated for the data sets \texttt{t.cell10} and \texttt{t.cell34} . This leads to a high number of selected edges for the Lasso and adaptive Lasso (see figures 4a and 4b), and consequently to a  high number of connected genes for these methods.

Figure 6 displays the percentage of positive ($>0$) correlations among the edges identified by the five methods for the six data sets. This proportion varies between 0.5 and 0.8. The results obtained using the five investigated methods seem much more consistent than the results on the number of identified edges.

We also compare the methods with respect to their stability. This is an important issue in order to assess the reliability of the methods \cite{Boulesteix09,Saeys07}: a good method is expected to yield a stable network in the sense that a slightly modified data set (for instance a subsample) does not lead to a completely different result. For data sets \texttt{ecoli1}, \texttt{ecoli2}, \texttt{t.cell10} and \texttt{t.cell34}, we draw subsamples by excluding $\approx 10\%$ of the observations and compute the network based on each subsample using the five methods successively. The number of considered subsamples is fixed to $R=10$ (only $R=9$ for the data set \texttt{ecoli2} that includes 9 observations). We do not analyze the data sets \texttt{ara} and \texttt{west}, because repeated experiments would be computationally too expensive.

For each candidate edge $i$, $n^{(1)} _i$ counts how often this edge is selected across the $R$ subsamples. Similarly, $n^{(0)} _i=R-n_i ^{(1)}$ denotes the number of times the $i$th edge is not selected. These frequencies are summarized using Fleiss' $\kappa$-score \cite{Fleiss71} which measures the degree of agreement among the $R$ subsamples of the data. The measure is defined as follows. We first compute the average proportion of assignments
\begin{eqnarray*}
p^{(l)}&=& \frac{1}{R \times \# \text{ edges}} \sum_{i=1} ^{\# \text{edges}} n_i ^{(l)},\,\,l=0,1\,.
\end{eqnarray*}
Further, the degree of agreement of the $R$ subsamples for the $i$th edge is measured as
\begin{eqnarray*}
P_i&=&\frac{1}{R(R+1)}\left[\sum_{l=0} ^1 \left(n_i ^{(l)} \right)^2 - R\right]
\end{eqnarray*}
Finally, with $\overline P$ denoting the average of the $P_i$'s and with $P_c= \left(p^{(0)}\right)^2 + \left(p^{(1)}\right)^2 $ denoting the agreement expected by chance, Fleiss $\kappa$ is defined as
\begin{eqnarray*}
\kappa&=& \frac{\overline P - P_c}{1-P_c}\,.
\end{eqnarray*}
The score is always $\leq$ 1, and the higher the value of $\kappa$, the more stable the methods are with respect to subsampling.

The $\kappa$-score of the methods is given in Table 4. As the absolute values are hard to compare between data sets, we also display the ranking on each data set. The shrinkage approach is the most stable, probably because it does not rely on additional subsampling in form of cross-validation splits. The regression based approaches are less stable, but among them, the degree of stability is comparable. In particular, in this experiment, we cannot see any difference between sparse and non-sparse approaches.

Finally, the considered methods differ quite dramatically with respect to their run-time. As an illustration, we compared the run-time on the \texttt{west} data set, which contains $3883$ genes. The approach based on shrinkage covariance estimation is by far the most efficient one ($\approx 2$ min), and all other methods scale within several \emph{hours}: PLS $\approx 7.5$ hours, ridge regression $\approx 10$ hours,  the Lasso $\approx 17$ hours, and the adaptive Lasso $\approx 3.5 $ \emph{days}. This can be seen as a major drawback of the methods relying on cross-validation schemes, especially the Lasso-based methods. While Ridge Regression and PLS allow a representation that only scales in the number of observations, Lasso and adaptive Lasso scale in the number of variables. Furthermore, adaptive Lasso requires nested cross-validation. Partial relief may be found in a parallel implementation. Alternatively, for high-dimensional data, one might consider to approximate the Lasso-based networks by first constructing a mildly sparse network without cross-validation (for example using the method described in \cite{Meinshausen0601}), and then to refine this network by running the (adaptive) Lasso with cross-validation.

\section{Discussion}

In this paper, we proposed and compared different methods to estimate partial correlation coefficients based on regularized regression techniques with applications to genetic networks.
In a simulation study, we assessed the performance of the considered methods in terms of estimation accuracy (MSE) and in terms of reverse engineering of the true underlying networked topology.  As a result, the investigated non-sparse methods (PLS, ridge regression, and the approach based on shrinkage covariance estimation that served as a reference method) were found to perform similarly.  It is to be noted that these methods have fdr-based significance testing  in common.  They are rather conservative with respect to the inclusion of edges when used with the standard fdr threshold 0.2.  The Lasso tends to produce too ``dense'' structures, while the adaptive Lasso compensates for that by selecting edges in a two-step approach, therefore leading to sparser graphs.  The latter two-stage approach is able to select relevant edges, even for small samples, while at the same time preventing to be too dense.  For denser networks, the performances of the five methods are very similar. On real world data, the behavior of the non-sparse methods is again similar, except that PLS is less conservative than ridge regression and the approach using a  shrinkage covariance estimator.  A remarkable difference with respect to the different data sets is the behavior of the Lasso and the adaptive Lasso on the t.cell data sets.  In contrast to the four other data sets, the t.cell data include replications, thus violating the assumption of independent samples.  Consequently, the (adaptive) Lasso does not handle the underlying data structure correctly, while empirical null modeling seems to account for the decreased ``effective'' sample size in an implicit way.\\
Note that all investigated methods require the specification of tuning parameters that need to be optimized based on the available data.
The choice of the model selection criterion itself strongly influences the results of the methods \cite{Bou2008}, especially for small $n$.
As an example, the model selection procedure introduces a substantial amount of variation for the Lasso and the adaptive Lasso.
In the real world study, we estimate the two graphs on two different random cross-validation splits, which leads to an overlap of only $88.4 \%$ on the \texttt{west} data, although the adaptive Lasso graph is defined as a subgraph of the Lasso graph.  Hence, tuning parameters should be given much attention in future research when new methods are developed.  Moreover, setting the parameters to fixed values without proper selection procedure (such as cross-validation) and just because they ``yield nice results''  is  an incorrect and biased strategy which may favor the proposed novel method.  Furthermore, from a computational point of view, a major strength of the shrinkage approach is that the optimal amount of regularization can be estimated from the data using an analytic formula, thus making time-consuming cross-validation procedures unnecessary.

We want to emphasize that there are interesting alternatives for the detection of significant edges that do not depend on sparsity penalties or testing based on local false discovery rates.  For instance, Reverter \& Chan \cite{Reverter0801} propose information theoretic measures for the reconstruction of gene co-expression networks.  The comparative performance of these methods and their connections to the approaches investigated above may be explored in future research.

Finally, the methods discussed in this paper can potentially be used for detecting causal interactions \cite{pellet07ida, Arnold0701}.  For instance, in the presence of longitudinal data, Arnold. et.al. \cite{Arnold0701} propose to identify the direction of interactions between variables by investigating partial correlations between time-shifted copies of the variables.  Amongst others, they propose to estimate these partial correlations using Lasso regression, but other regression methods might be promising alternatives.

\section{Conclusion}
We briefly summarize our findings.

\emph{Performance:} In the simulation, the investigated non-sparse regression methods, i.e. Ridge Regression and Partial Least Squares, exhibit rather conservative behavior when combined with (local) false discovery rate multiple testing in order to decide whether or not an edge is present in the network. For networks with higher densities, the difference in performance of the methods decreases. Both sparse and non-sparse methods can deal with cluster topologies in the network.

For PLS, we observe both a high MSE in the simulations and a high percentage of selected edges in some of the real data. In our opinion, this is an indication that PLS itself might not be too well-suited for the reconstruction of networks. The reasons are that  PLS is not sparse by design, and that it does not shrink arbitrarily close to zero. Therefore, we suggest to incorporate sparse versions of PLS instead in future research.

On six real data sets, we also clearly distinguish the results obtained using the non-sparse methods and those obtained using the sparse methods where specification of the regularization parameter automatically means model selection.
For data that violate the assumption of uncorrelated observations (due to replications), the Lasso and the adaptive Lasso yield very complex structures, indicating that they might not be suited under these conditions.

\emph{Stability:} We compared the stability of the methods under two aspects. All regression-based methods are less stable than the shrinkage approach over different subsamples of the data, and within the regression-based approaches, there is no clear difference between sparse and non-sparse methods. However, the two sparse regression methods seem to be unstable with respect to violations of the i.i.d assumption of the samples.

\emph{Runtime:} The computational load for the Lasso and in particular for the adaptive Lasso is considerable. For very high-dimensional data, this can constitute a severe limitation. The runtime might be decreased by applying parallel computation techniques or by preselecting a coarse network topology that does not rely on cross-validation. While PLS and Ridge Regression are slower than shrinkage, both of them are fairly fast to compute, as they allow a kernel representation, i.e. most of the computation scales in the number of samples and not in the number of variables.

\section*{Available Software}

The regularized estimation of partial correlations and the construction of gene association networks with (adaptive) Lasso, ridge regression and
PLS are implemented in the \texttt{R} package \texttt{parcor} \cite{parcor} which is available from the CRAN repository \url{http://cran.r-project.org/}. The package relies heavily on the \texttt{lars} package \cite{lars}.  For assigning statistical significane to the edges in the network, we use the \texttt{fdrtool} package \cite{Strimmer0803}.  An executable sheet for the simulations can be downloaded\footnote{\url{http://ml.cs.tu-berlin.de/~nkraemer/software.html}}.

\section*{Authors' contribution}

NK and ALB initiated the study.  NK wrote the initial version of the  manuscript.  JS and NK implemented the R package, NK and ALB performed the analyses. All authors contributed to the concept and to the manuscript.

\section*{Acknowledgments}

This work was supported in part by the BMBF grant FKZ 01-IS07007A (ReMind), and the FP7-ICT Programme of the European Community, under the PASCAL2 Network of Excellence, ICT-216886.  Financial support from DSM Nutritional Products Ltd. (Basel, Switzerland) is gratefully acknowledged.  We thank Lukas Meier and Mikio L. Braun for constructive comments on model selection, and Animesh Acharjee for helpful feedback on the R package ``parcor''.


\providecommand{\bysame}{\leavevmode\hbox to3em{\hrulefill}\thinspace}
\providecommand{\MR}{\relax\ifhmode\unskip\space\fi MR }
\providecommand{\MRhref}[2]{%
  \href{http://www.ams.org/mathscinet-getitem?mr=#1}{#2}
}
\providecommand{\href}[2]{#2}


\newpage
\appendix
\section{Figures}
  \subsection*{Figure 1a - MSE, number of edges, power and TDR for a density of $0.05$}
      Mean squared error, number of selected edges,
power and  true discovery rate (TDR) for the various methods PLS, ridge regression,
the approach based on shrinkage covariance estimation, Lasso, and adaptive Lasso.

\begin{figure}[htb]
{\par\centering\resizebox*{16cm}{16cm}{\rotatebox{0}{{\includegraphics{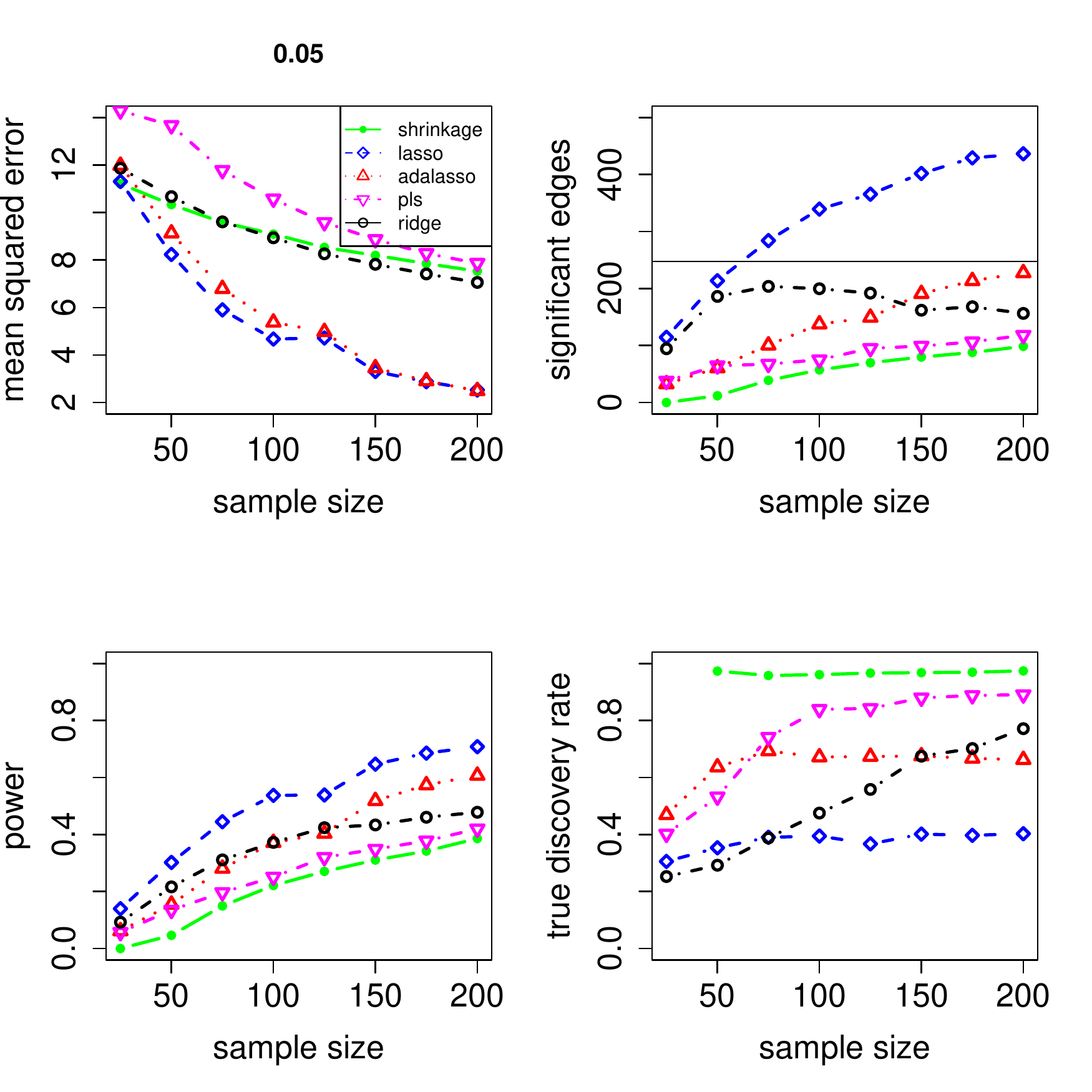}}}}\par}
\end{figure}
\newpage
\subsection*{Figure 1b - MSE, number of edges, power and TDR for a density of $0.10$}
      Mean squared error, number of selected edges,
power and  true discovery rate (TDR) for the various methods PLS, ridge regression,
the approach based on shrinkage covariance estimation, Lasso, and adaptive Lasso.

\begin{figure}[htb]
{\par\centering\resizebox*{16cm}{16cm}{\rotatebox{0}{{\includegraphics{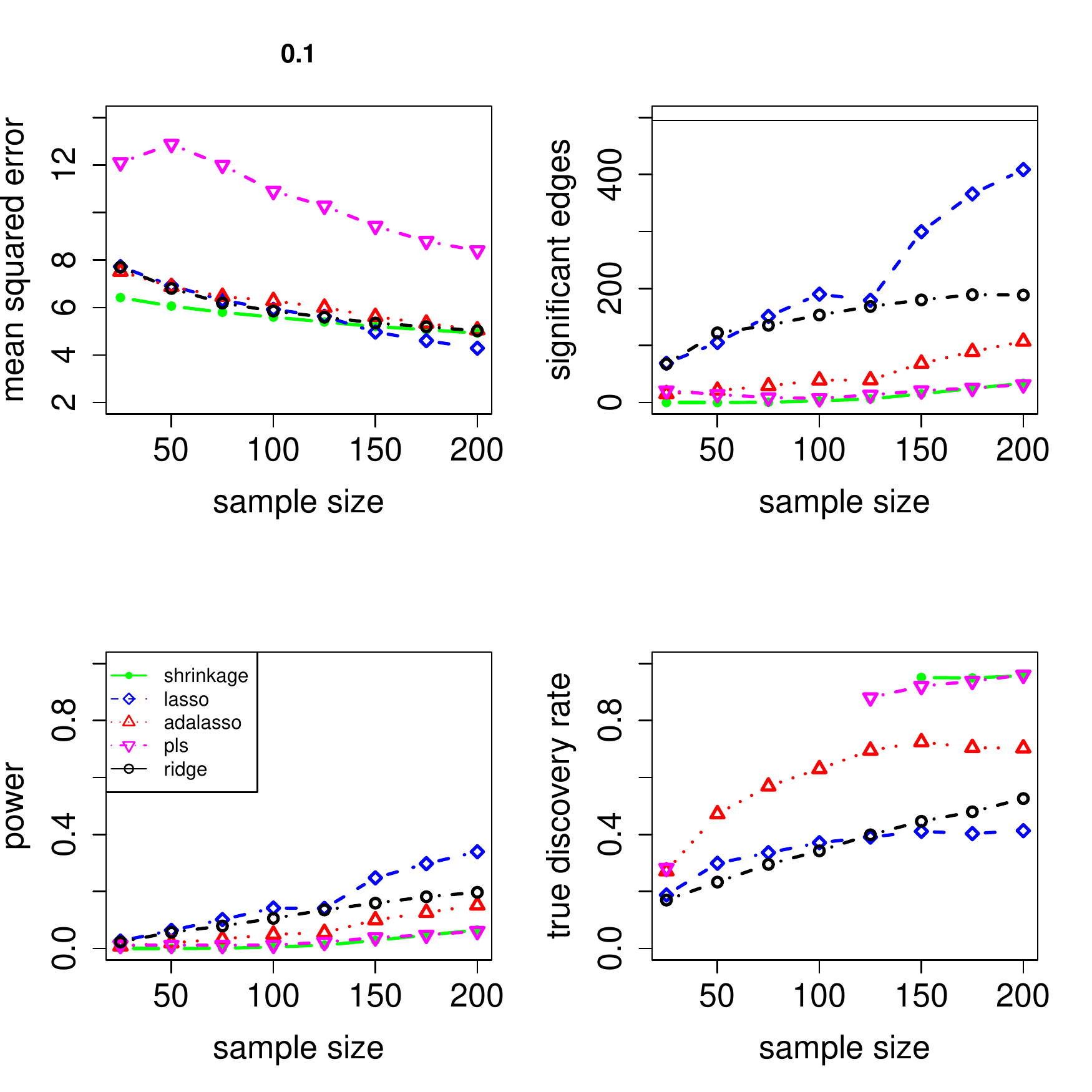}}}}\par}
\end{figure}
\newpage
\subsection*{Figure 1c - MSE, number of edges, power and TDR for a density of $0.15$}
      Mean squared error, number of selected edges,
power and  true discovery rate (TDR) for the various methods PLS, ridge regression,
the approach based on shrinkage covariance estimation, Lasso, and adaptive Lasso.

\begin{figure}[htb]
{\par\centering\resizebox*{16cm}{16cm}{\rotatebox{0}{{\includegraphics{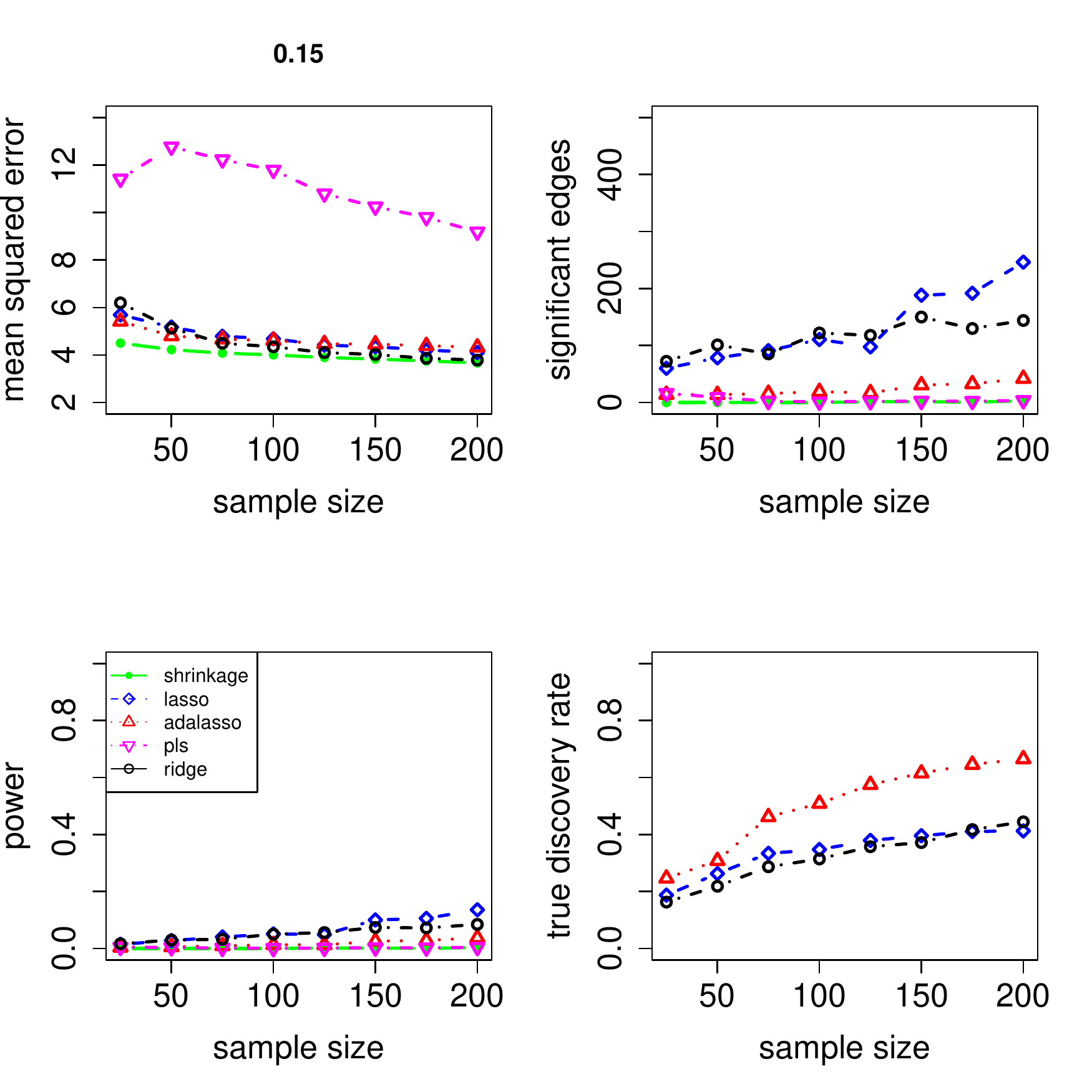}}}}\par}
\end{figure}
\newpage
\subsection*{Figure 1d - MSE, number of edges, power and TDR for a density of $0.20$}
      Mean squared error, number of selected edges,
power and  true discovery rate (TDR) for the various methods PLS, ridge regression,
the approach based on shrinkage covariance estimation, Lasso, and adaptive Lasso.

\begin{figure}[htb]
{\par\centering\resizebox*{16cm}{16cm}{\rotatebox{0}{{\includegraphics{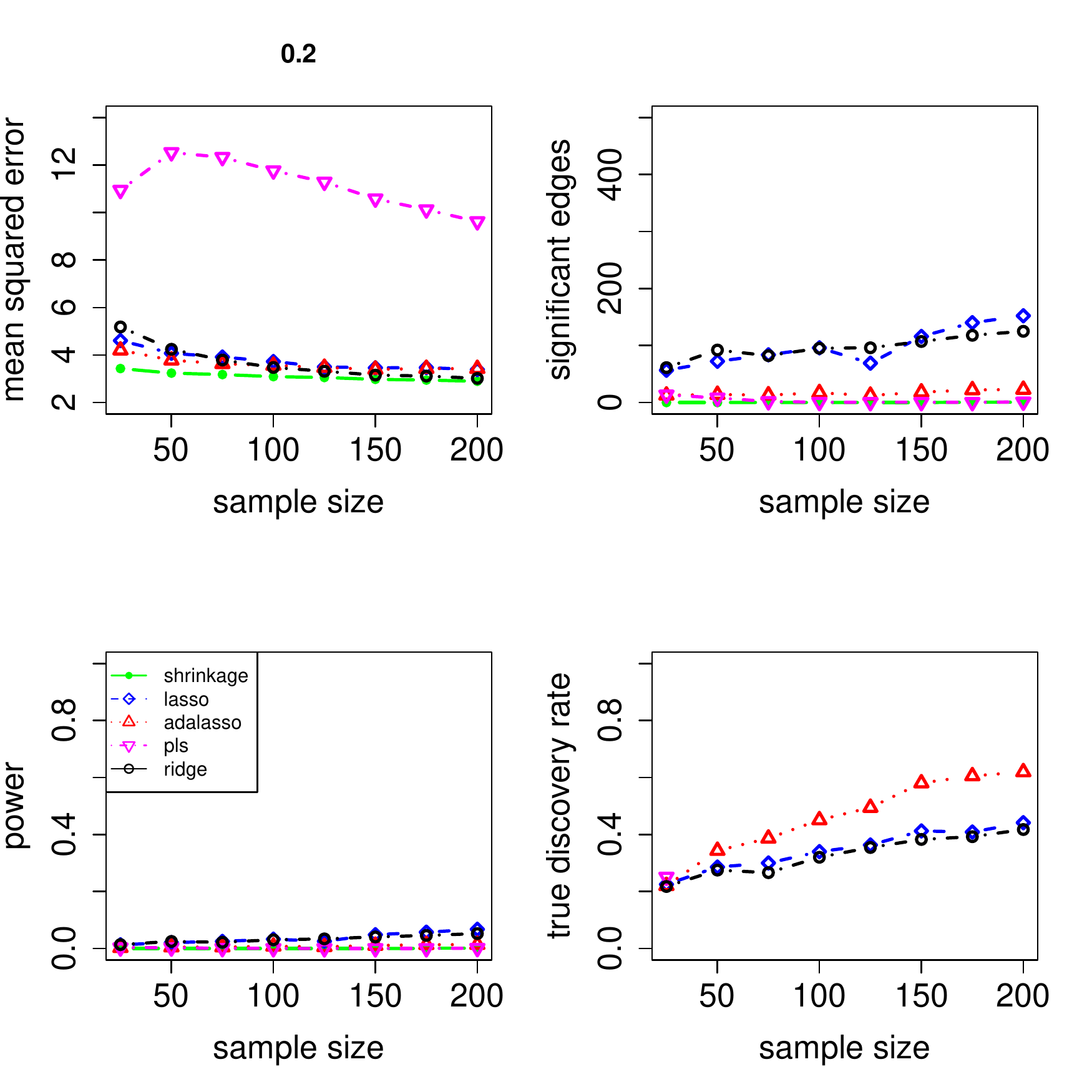}}}}\par}
\end{figure}
\newpage
\subsection*{Figure 1e - MSE, number of edges, power and TDR for a density of $0.25$}
      Mean squared error, number of selected edges,
power and  true discovery rate (TDR) for the various methods PLS, ridge regression,
the approach based on shrinkage covariance estimation, Lasso, and adaptive Lasso.

\begin{figure}[htb]
{\par\centering\resizebox*{16cm}{16cm}{\rotatebox{0}{{\includegraphics{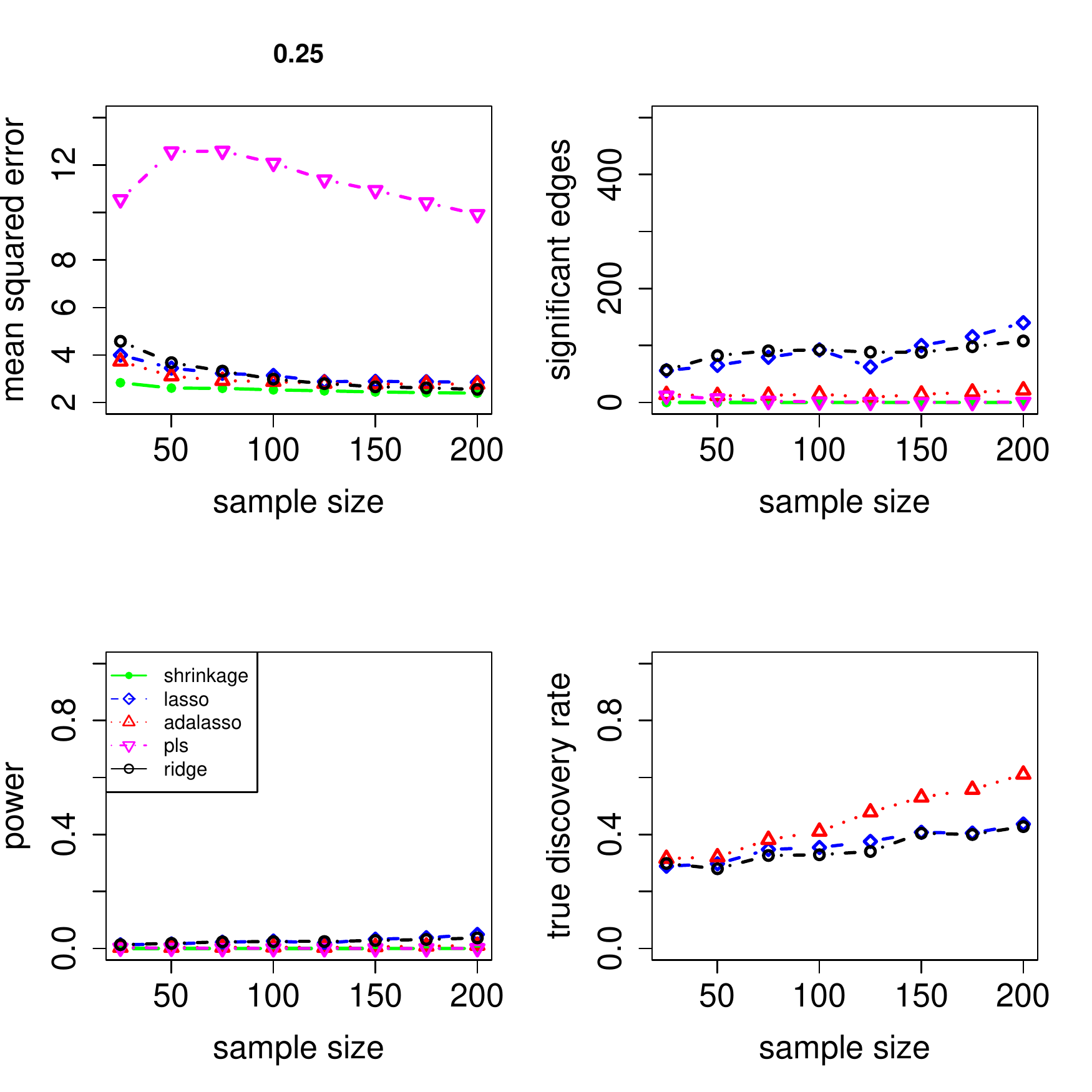}}}}\par}
\end{figure}
\newpage

\subsection*{Figure 2 - Runtime of the respective methods}

\begin{figure}[htb]
{\par\centering\resizebox*{7cm}{7cm}{\rotatebox{0}{{\includegraphics{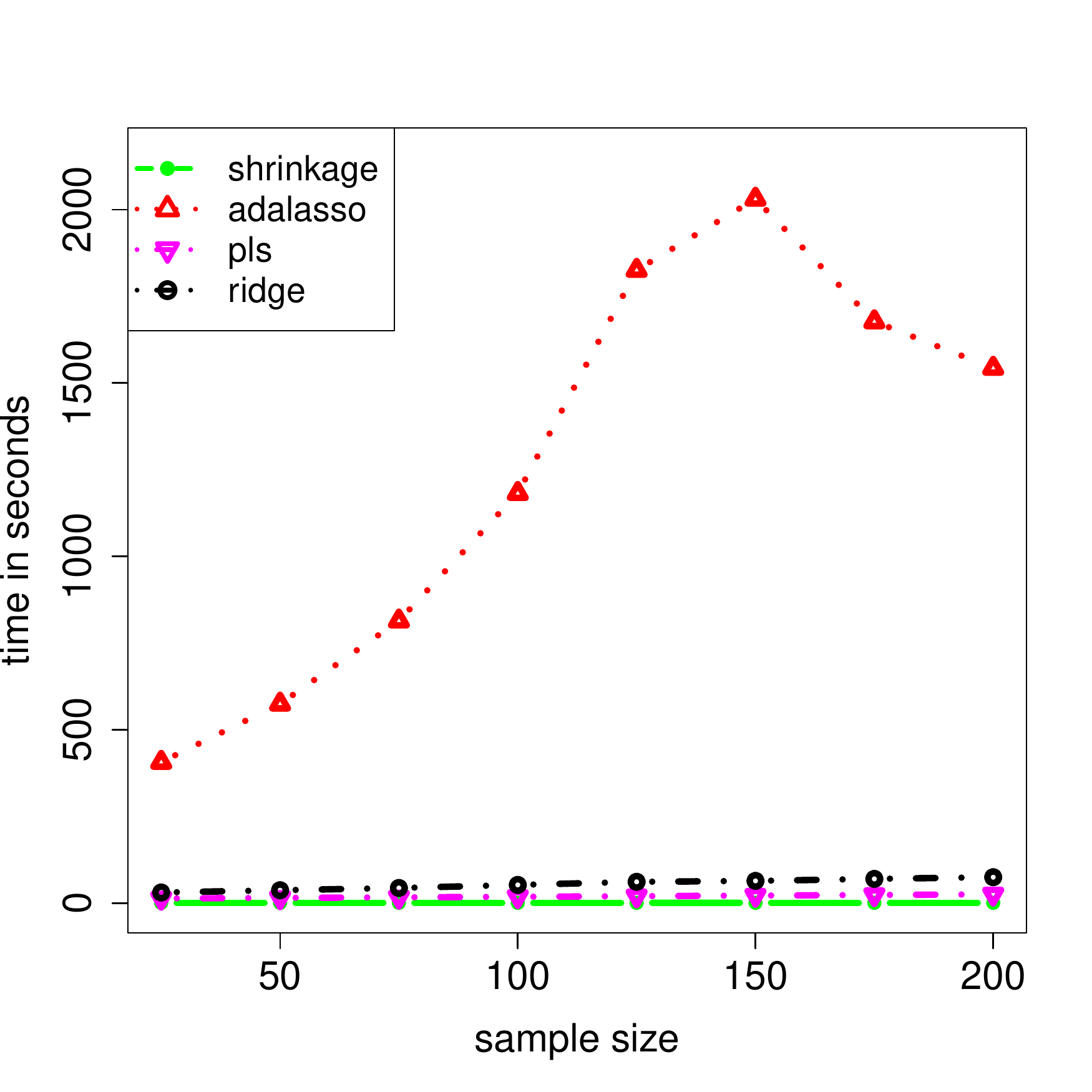}}}}\par}
\end{figure}

\begin{figure}[htb]
{\par\centering\resizebox*{7cm}{7cm}{\rotatebox{0}{{\includegraphics{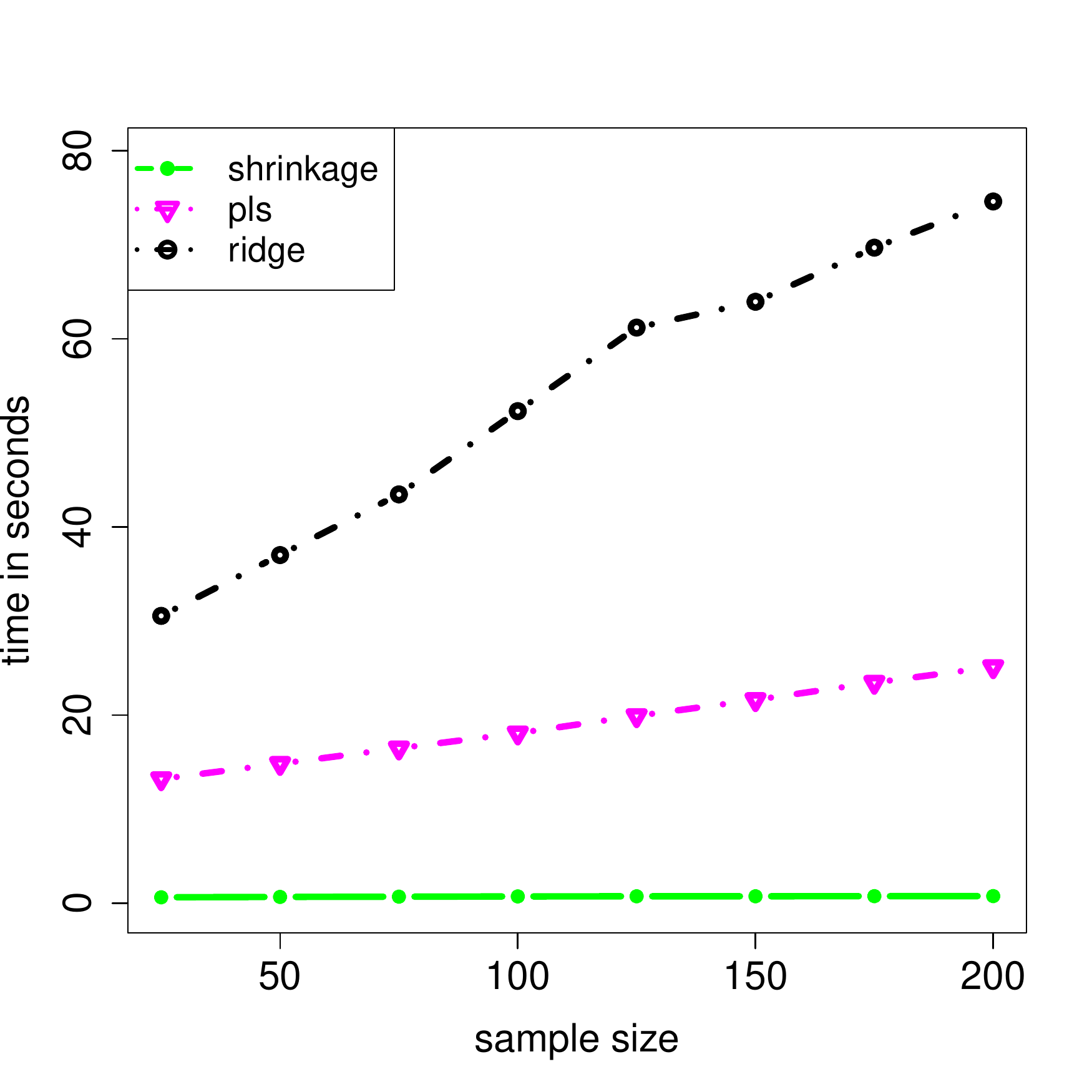}}}}\par}
\end{figure}
\newpage

\subsection*{Figure 3a - Network topology: 1 cluster}

\begin{figure}[htb]
{\par\centering\resizebox*{16cm}{16cm}{\rotatebox{0}{{\includegraphics{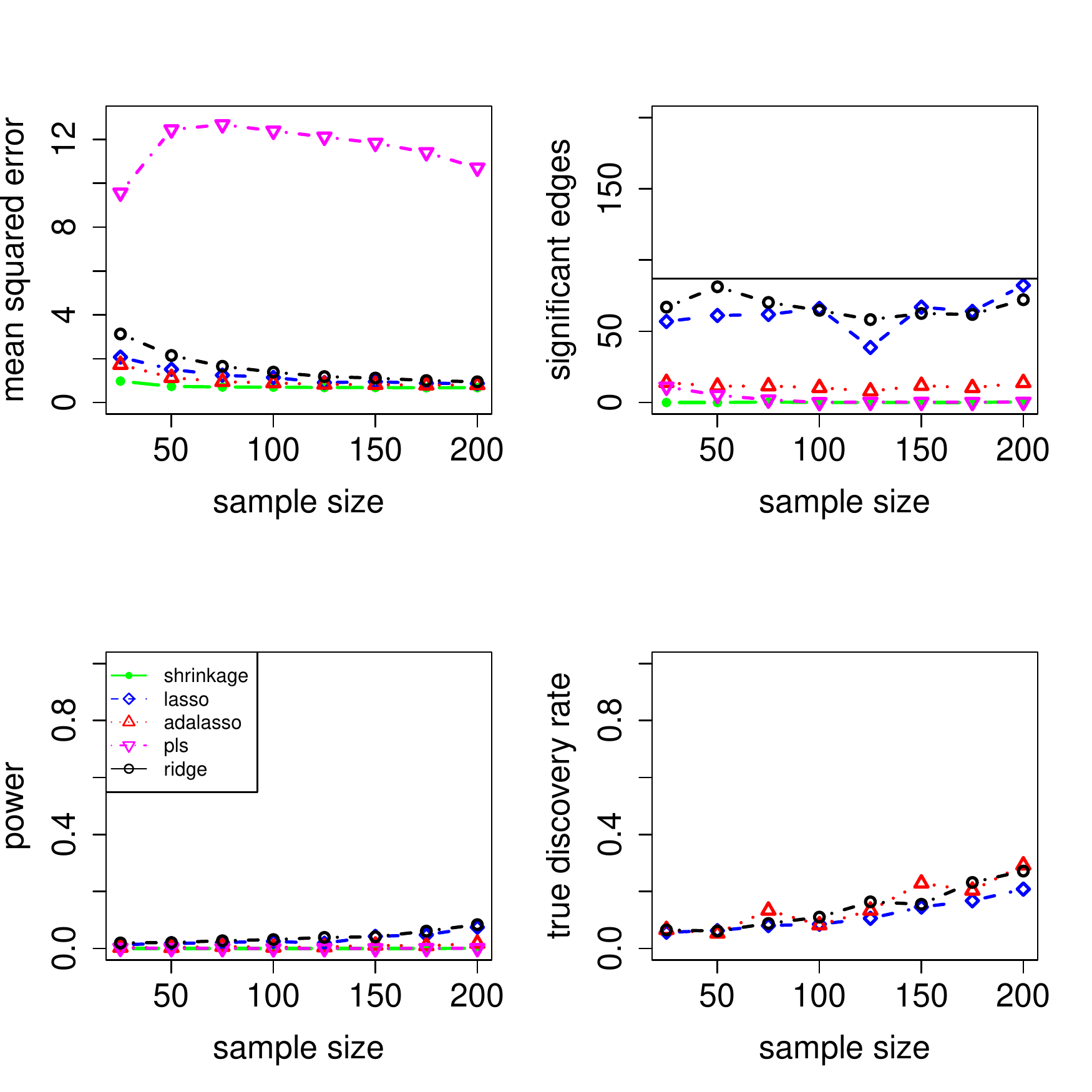}}}}\par}
\end{figure}

\newpage

\subsection*{Figure 3b - Network topology: 2 clusters}

\begin{figure}[htb]
{\par\centering\resizebox*{16cm}{16cm}{\rotatebox{0}{{\includegraphics{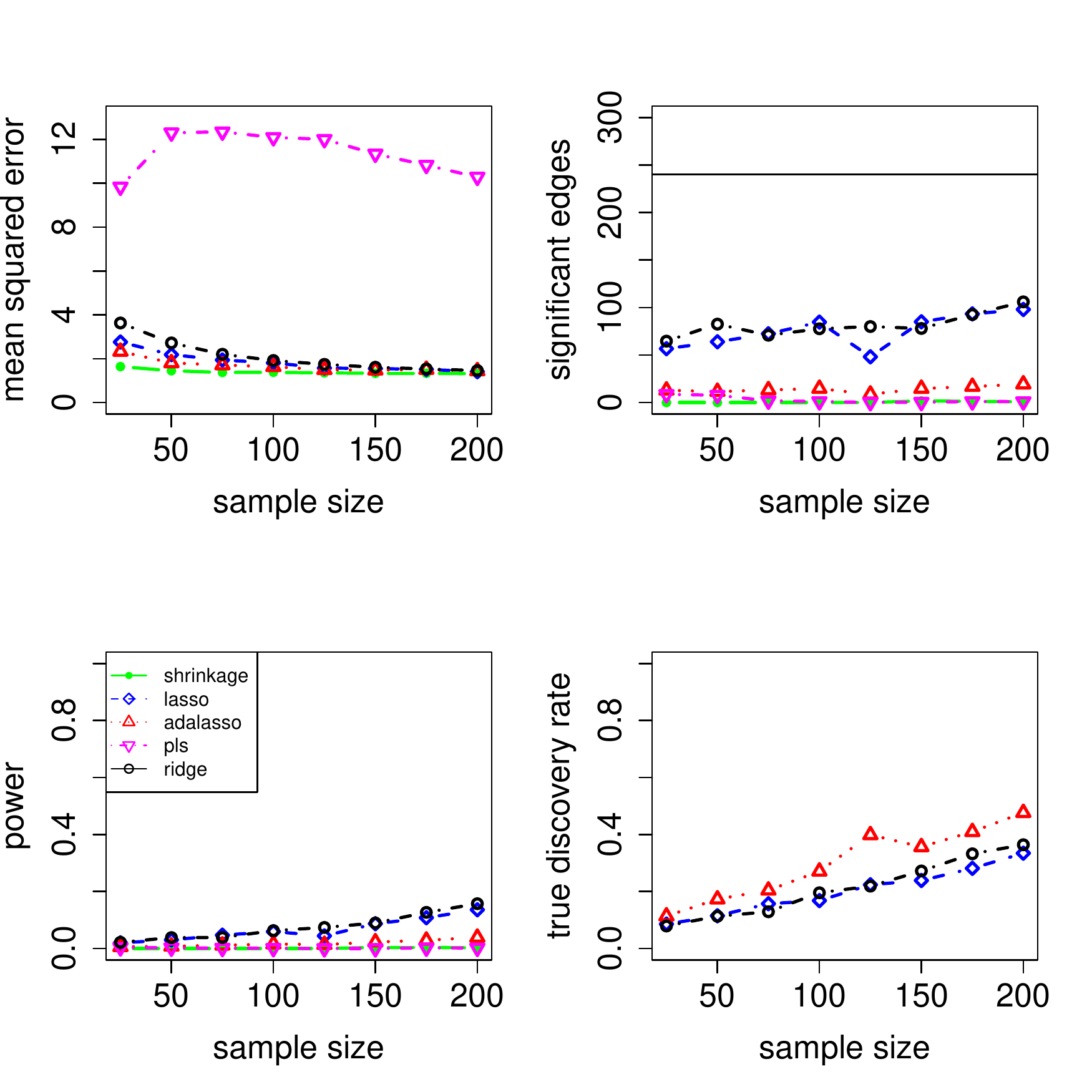}}}}\par}
\end{figure}

\newpage

\subsection*{Figure 3c - Network topology: 3 clusters}

\begin{figure}[htb]
{\par\centering\resizebox*{16cm}{16cm}{\rotatebox{0}{{\includegraphics{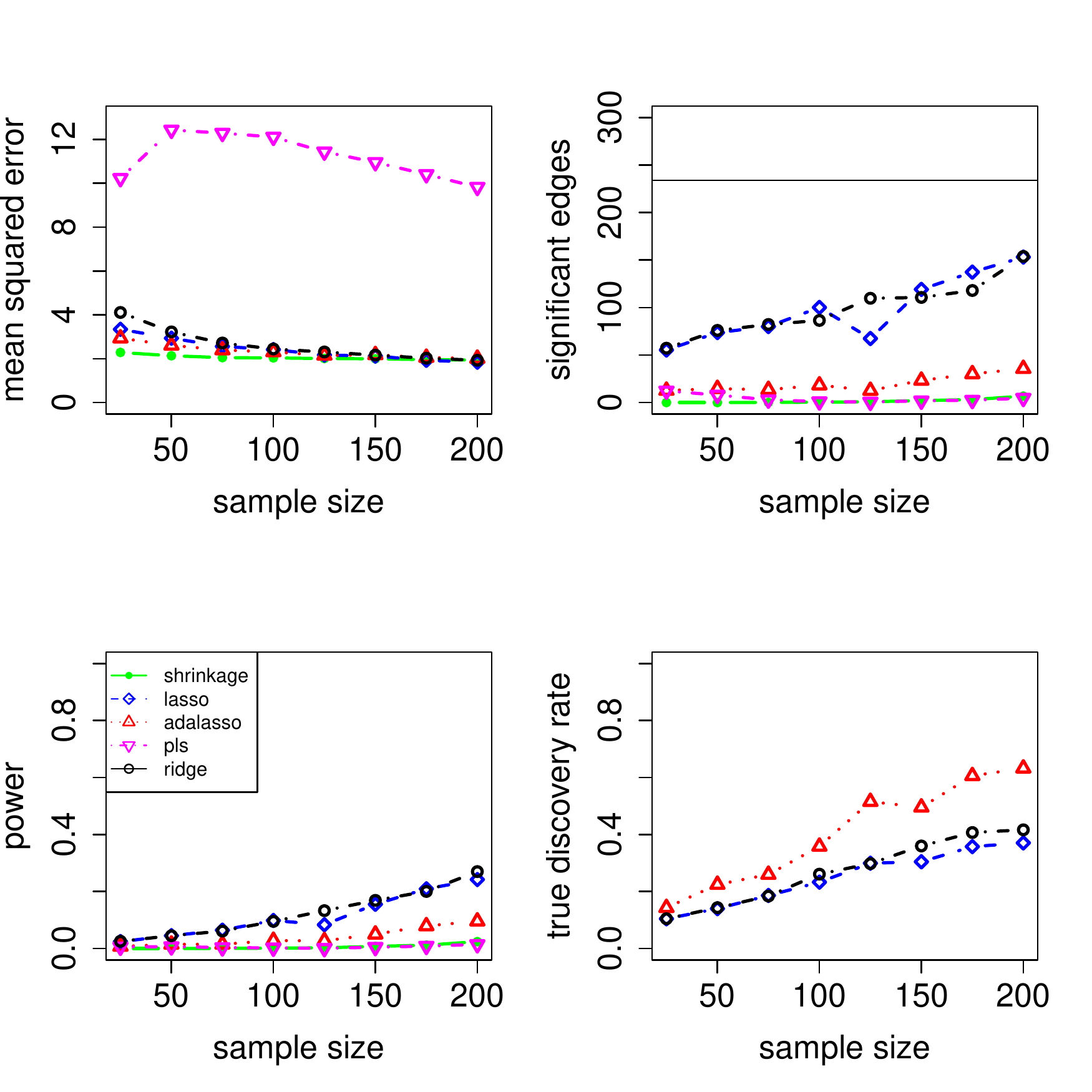}}}}\par}
\end{figure}

\newpage

\subsection*{Figure 3d - Network topology: 3 stars}

\begin{figure}[htb]
{\par\centering\resizebox*{16cm}{16cm}{\rotatebox{0}{{\includegraphics{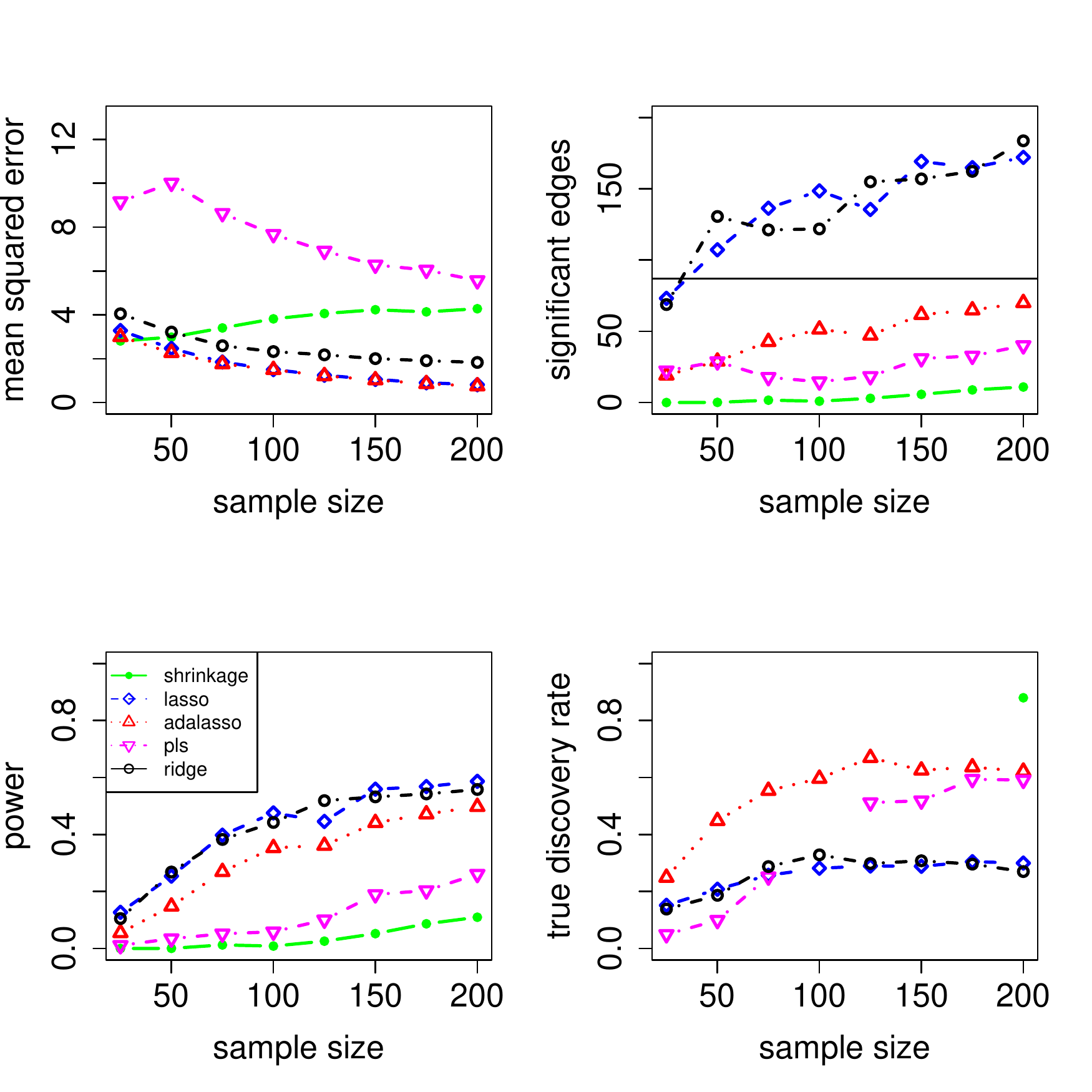}}}}\par}
\end{figure}

\newpage

\subsection*{Figure 4a - Proportion of selected edges}

\begin{figure}[htb]
{\par\centering\resizebox*{16cm}{16cm}{\rotatebox{0}{{\includegraphics{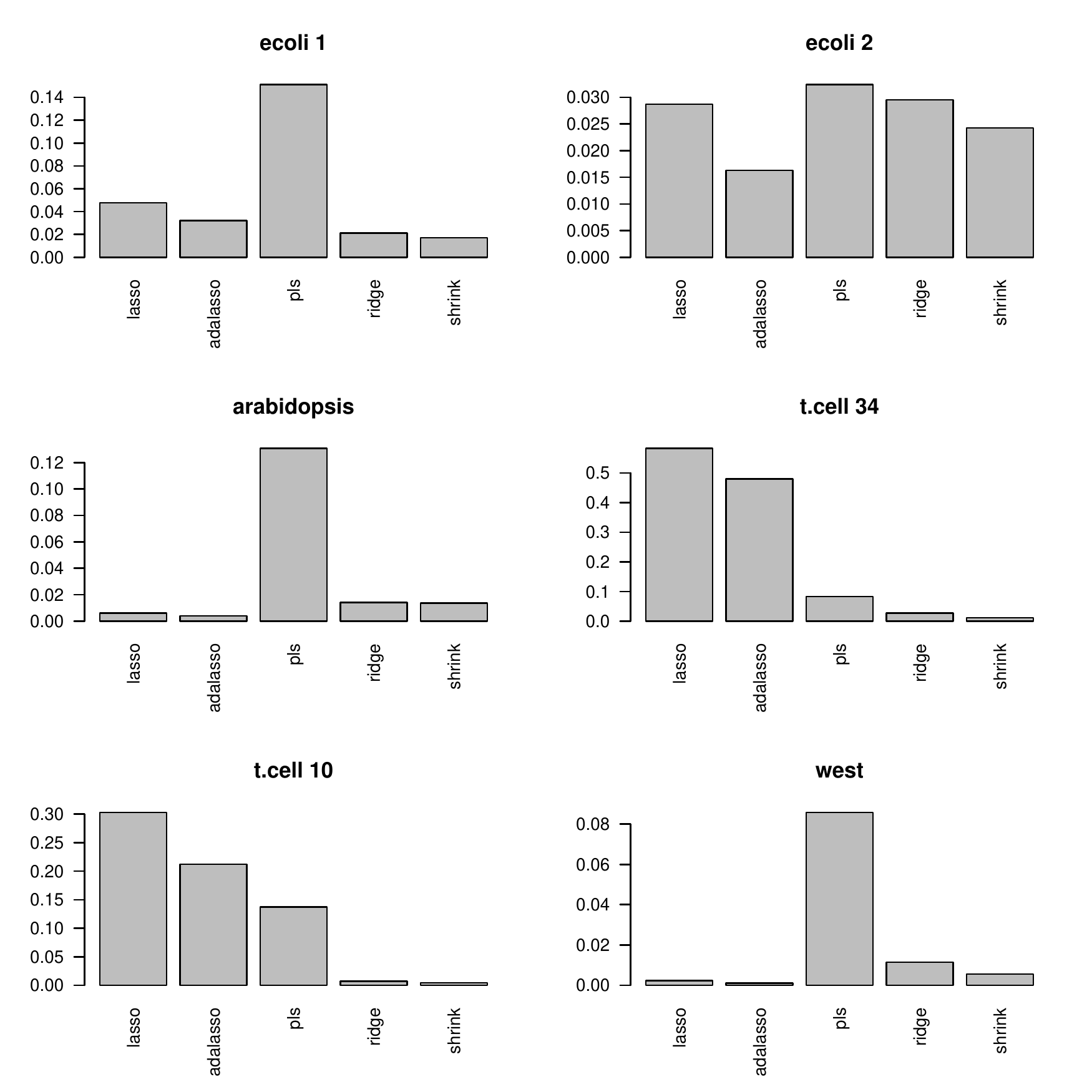}}}}\par}
\end{figure}
\newpage
\subsection*{Figure 4b - Proportion of selected edges without PLS}

\begin{figure}[htb]
{\par\centering\resizebox*{16cm}{16cm}{\rotatebox{0}{{\includegraphics{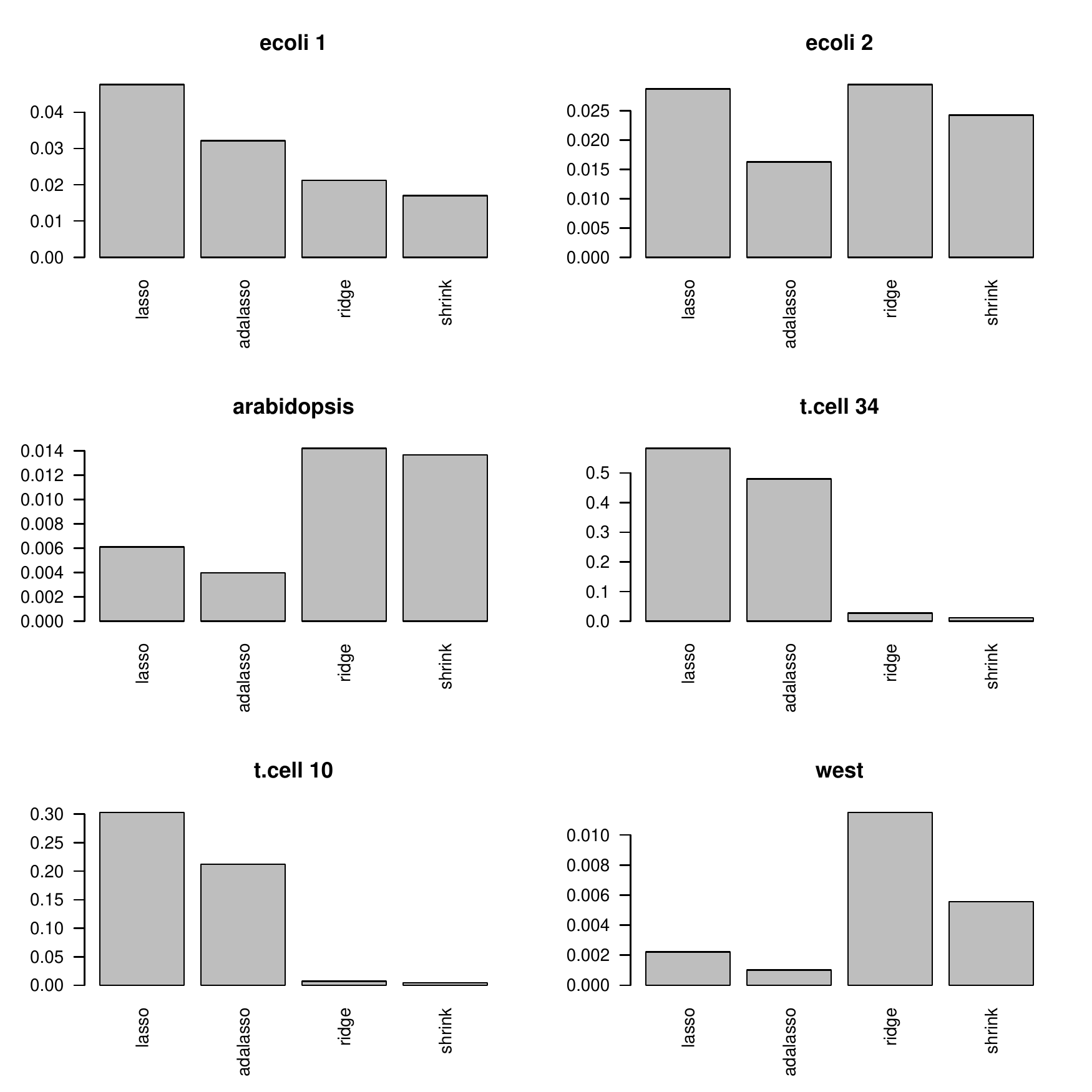}}}}\par}
\end{figure}
\newpage
\subsection*{Figure 5a - Connectivity: Proportion of connected genes}

\begin{figure}[htb]
{\par\centering\resizebox*{16cm}{16cm}{\rotatebox{0}{{\includegraphics{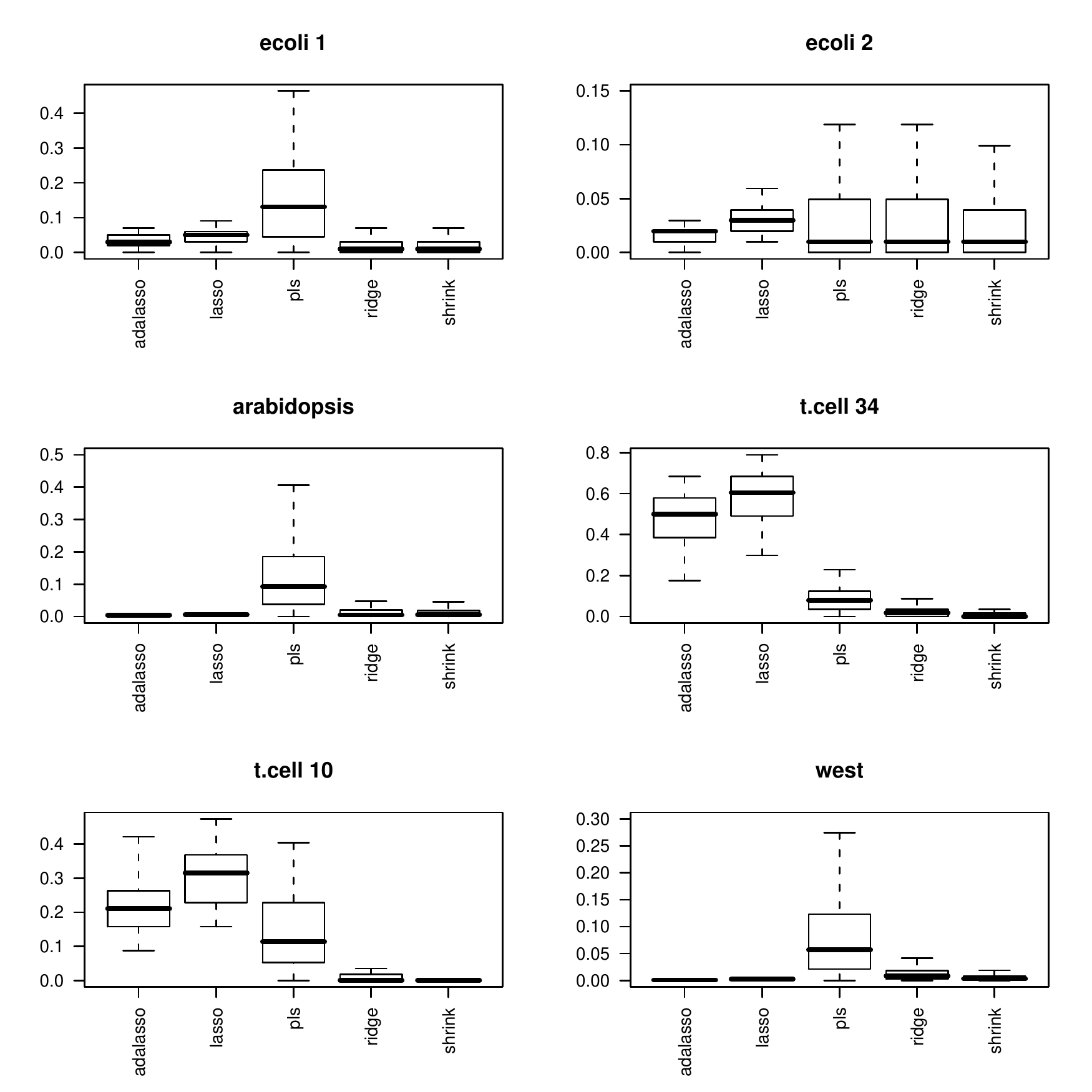}}}}\par}
\end{figure}

\newpage

\subsection*{Figure 5b - Connectivity: Proportion of connected genes without PLS}

\begin{figure}[htb]
{\par\centering\resizebox*{16cm}{16cm}{\rotatebox{0}{{\includegraphics{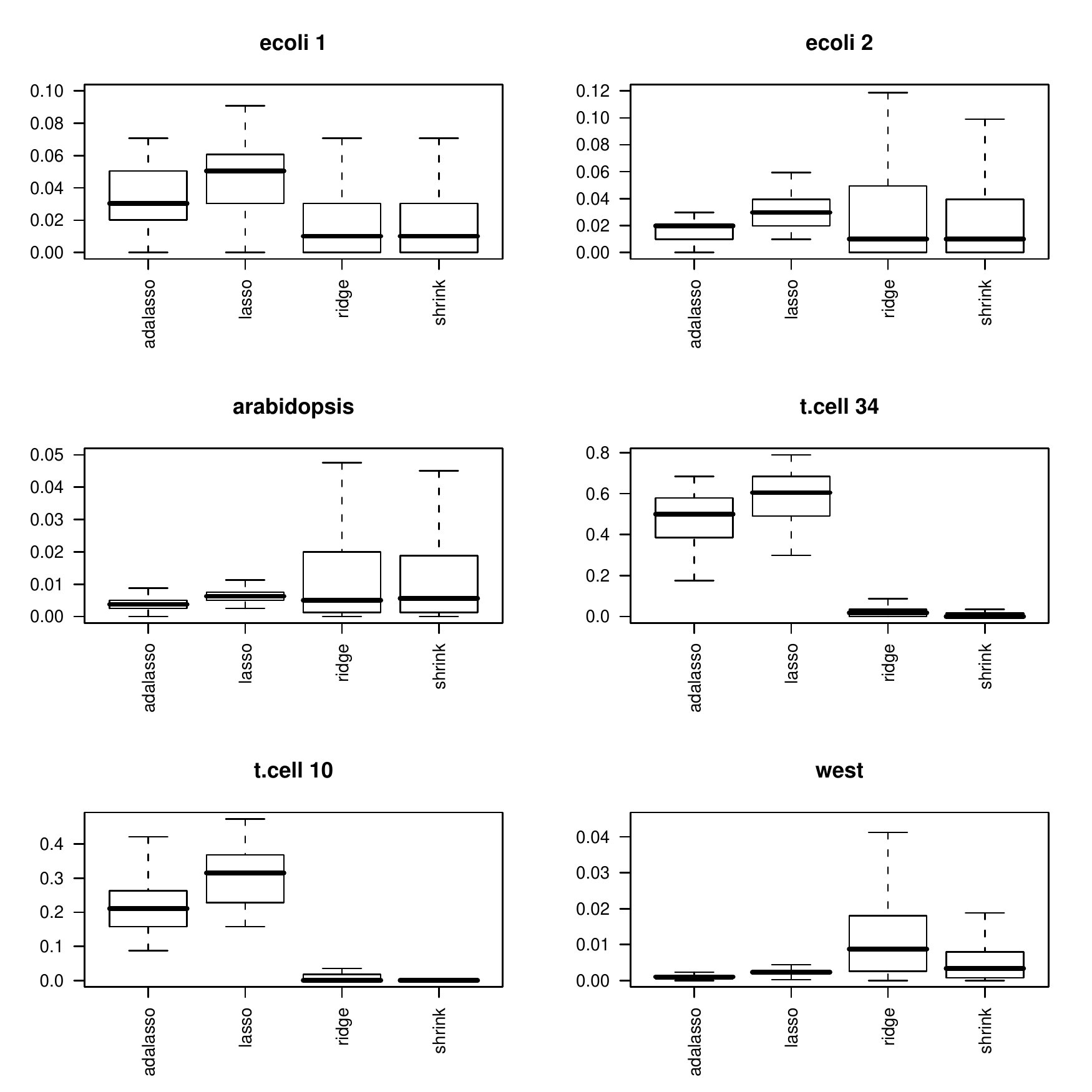}}}}\par}
\end{figure}
\newpage
\subsection*{Figure 6 - Percentage of positive correlations}

\begin{figure}[htb]
{\par\centering\resizebox*{16cm}{16cm}{\rotatebox{0}{{\includegraphics{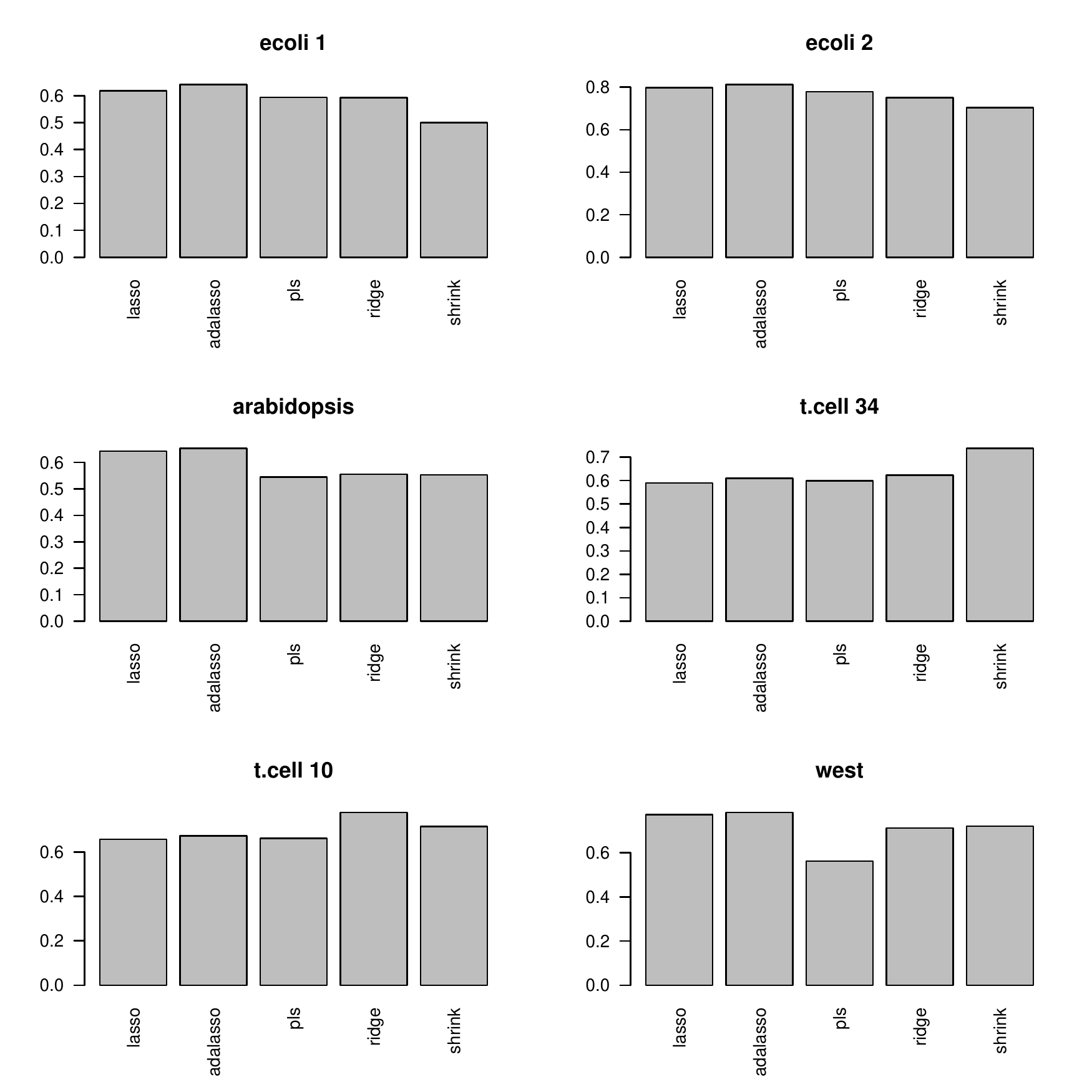}}}}\par}
\end{figure}
\newpage

\section{Tables}

\subsection*{Table 1 - Overview of the methods}
Characteristics of the five methods considered in this study: (1) shrinkage estimator of the covariance (shrink), (2) PLS regression (pls), (3) Ridge Regression (ridge), (4) Lasso regression (lasso), (5) Adaptive Lasso (adalasso). {\bf 2nd column:} Type of the method (inversion of a regularized estimate of the covariance matrix or regression-type method). {\bf 3rd column:} Parameter determining the amount of regularization. {\bf 4th column:} Method used to choose this(these) parameter(s). {\bf 5th column:} Method used to decide whether two genes should be edge-connected.
\vspace{0.5cm}
\begin{center}
\begin{tabular}{lcccc}
\hline
Method& type & parameter(s) & choice & edge if\\ \hline
shrink & \begin{tabular}{c}regularized estimation\\ of the covariance\end{tabular} & shrinkage intensity $\lambda$ & analytic & fdr$<0.2$ \\
pls & regression & number of components $m$ & CV & fdr$<0.2$ \\
ridge  & regression & penalty $\lambda$ & CV & fdr$<0.2$ \\
lasso & regression & penalty $\lambda$ & CV & $\widehat \rho_{ij}\neq 0$ \\
adalasso & regression & penalty $\lambda$ ($ \times 2$) & nested CV ($\times 2$)& $\widehat \rho_{ij}\neq 0$\\
\hline
\end{tabular}
\end{center}

\subsection*{Table 2 - Size of the data sets}
\vspace{0.5cm}
\begin{center}
\begin{tabular}{lcccccl}
\hline
data set& arrays &genes & time series &\begin{tabular}{c}repeated\\measurements\end{tabular}&\begin{tabular}{c}size of\\ full graph\end{tabular} & Availability\\
\hline
\texttt{ecoli1}& 23  &100 & yes& no&$4\,950$ & \texttt{R} package \texttt{plsgenomics} \cite{plsgenomics}\\
\texttt{ecoli2}& 9& 102& yes&no&$5\,151$ & \texttt{R} package \texttt{GeneNet} \cite{Schaefer0601}\\
\texttt{ara}&22&800& yes&no&$319\,600$ & \texttt{R} package \texttt{GeneNet} \\
\texttt{t.cell10} &100&58& yes&yes&$1\,653$ &\texttt{R} package \texttt{longitudinal} \cite{longitudinal}\\
\texttt{t.cell34} &340&58& yes&yes&$1\,653$ &\texttt{R} package \texttt{longitudinal}\\
\texttt{west}&49&3883&no&no&$7\,536\,903$ & {\small{\url{http://strimmerlab.org/data.html}}}\\
\hline
\end{tabular}
\end{center}
\newpage

\subsection*{Table 3 - Overlap of the estimated graphs}
Example: On the \texttt{ecoli1} data set, $68,6\%$ of the edges
found by Ridge Regression are also found by PLS. For baseline comparison, the numbers in {\emph{italics}} show the percentage of selected edges for the respective methods.

\vspace{0.5cm}
\begin{center}
\begin{tabular}{llccccc}
\hline
data set&&pls &ridge& lasso& adalasso& shrink\\
  \hline
&pls & 1.000 & 0.096 & 0.156 & 0.127 & 0.045 \\
 & ridge & 0.686 & 1.000 & 0.600 & 0.457 & 0.390 \\
 \texttt{ecoli1}& lasso & 0.496 & 0.267 & 1.000 & 0.581 & 0.165 \\
 & adalasso & 0.597 & 0.302 & 0.862 & 1.000 & 0.189 \\
 & shrink & 0.405 & 0.488 & 0.464 & 0.357 & 1.000 \\
 &\emph{\% selected}& \it{0.162}& \it{0.018}&\it{0.052}&\it{0.036}&\it{0.017}\\
   \hline
&pls & 1.000 & 0.593 & 0.263 & 0.156 & 0.305 \\
&  ridge & 0.651 & 1.000 & 0.309 & 0.197 & 0.388 \\
\texttt{ecoli2} & lasso & 0.297 & 0.318 & 1.000 & 0.520 & 0.311 \\
 & adalasso & 0.310 & 0.357 & 0.917 & 1.000 & 0.381 \\
 & shrink & 0.408 & 0.472 & 0.368 & 0.256 & 1.000 \\
 &\emph{\% selected}& \it{0.032}& \it{0.030}&\it{0.029}&\it{0.020}&\it{0.024}\\
   \hline
&pls & 1.000 & 0.064 & 0.025 & 0.017 & 0.035 \\
 & ridge & 0.590 & 1.000 & 0.151 & 0.108 & 0.377 \\
\texttt{ara} & lasso & 0.535 & 0.352 & 1.000 & 0.579 & 0.361 \\
 & adalasso & 0.556 & 0.386 & 0.887 & 1.000 & 0.409 \\
 & shrink & 0.335 & 0.392 & 0.161 & 0.119 & 1.000 \\
 &\emph{\% selected}& \it{0.126}& \it{0.018}&\it{0.006}&\it{0.004}&\it{0.014}\\
   \hline
&pls & 1.000 & 0.314 & 0.993 & 0.985 & 0.131 \\
 & ridge & 0.956 & 1.000 & 1.000 & 1.000 & 0.422 \\
\texttt{t.cell10}  &lasso & 0.141 & 0.047 & 1.000 & 0.795 & 0.020 \\
  &adalasso & 0.170 & 0.057 & 0.965 & 1.000 & 0.024 \\
  &shrink & 0.947 & 1.000 & 1.000 & 1.000 & 1.000 \\
  &\emph{\% selected}& \it{0.109}& \it{0.027}&\it{0.575}&\it{0.417}&\it{0.011}\\
   \hline
&pls & 1.000 & 0.053 & 0.762 & 0.670 & 0.031 \\
 & ridge & 1.000 & 1.000 & 1.000 & 1.000 & 0.583 \\
\texttt{t.cell34} & lasso & 0.345 & 0.024 & 1.000 & 0.643 & 0.014 \\
 & adalasso & 0.433 & 0.034 & 0.917 & 1.000 & 0.020 \\
 & shrink & 1.000 & 1.000 & 1.000 & 1.000 & 1.000 \\
 &\emph{\% selected}& \it{0.134}& \it{0.005}&\it{0.284}&\it{0.221}&\it{0.004}\\
   \hline
& pls & 1.000 & 0.089 & 0.017 & 0.008 & 0.041 \\
 & ridge & 0.667 & 1.000 & 0.118 & 0.062 & 0.236 \\
\texttt{west} & lasso & 0.643 & 0.611 & 1.000 & 0.407 & 0.404 \\
 & adalasso & 0.673 & 0.694 & 0.884 & 1.000 & 0.458 \\
 & shrink & 0.632 & 0.488 & 0.161 & 0.084 & 1.000 \\
 &\emph{\% selected}& \it{0.086}& \it{0.011}&\it{0.002}&\it{0.001}&\it{0.006}\\
   \hline
\end{tabular}
\end{center}

\newpage

\subsection*{Table 4 - Stability of the Methods}
For the data sets \texttt{ecoli1}, \texttt{ecoli2}, \texttt{t.cell10}, \texttt{t.cell34} and \texttt{west}, we display Fleiss' kappa. The quantity is always $\leq 1$, and the higher the value, the more stable is the method.

\vspace*{1cm}

\begin{center}
\begin{tabular}{lrccccc}
\hline
data set& measure &pls&ridge & lasso & adalasso&shrink\\
\hline
\texttt{ecoli1}& $\kappa$& 0.630& 0.510& 0.550& 0.550& 0.593\\
& ranking of $\kappa$& 1 &5&3.5&3.5&2\\
\hline
\texttt{ecoli2}&$\kappa$& 0.242& 0.280 &0.469 &0.450& 0.486\\
&ranking of $\kappa$&5&4&2&3&1\\
\hline
\texttt{t.cell10}& $\kappa$&0.656& 0.797 &0.670  &0.674 &0.742\\
&ranking of $\kappa$&5&1&4&3&2\\
\hline
\texttt{t.cell34}&$\kappa$&0.655& 0.555& 0.625& 0.619 &0.702\\
&ranking of $\kappa$&2& 5&3&4&1\\
\hline
&mean ranking of $\kappa$&3.25&3.75&3.125&3.375&1.5\\
\hline

\end{tabular}
\end{center}
\newpage
\section{Supplementary Material}
\subsection*{Figure 2a - ROC curves for a density of $0.05$}
      ROC curves obtained by varying the fdr-threshold for PLS, Ridge Regression and Shrinkage. The sensitivity and specificity of Lasso and Adaptive Lasso are represented by a point.

\begin{figure}[htb]
{\par\centering\resizebox*{16cm}{16cm}{\rotatebox{0}{{\includegraphics{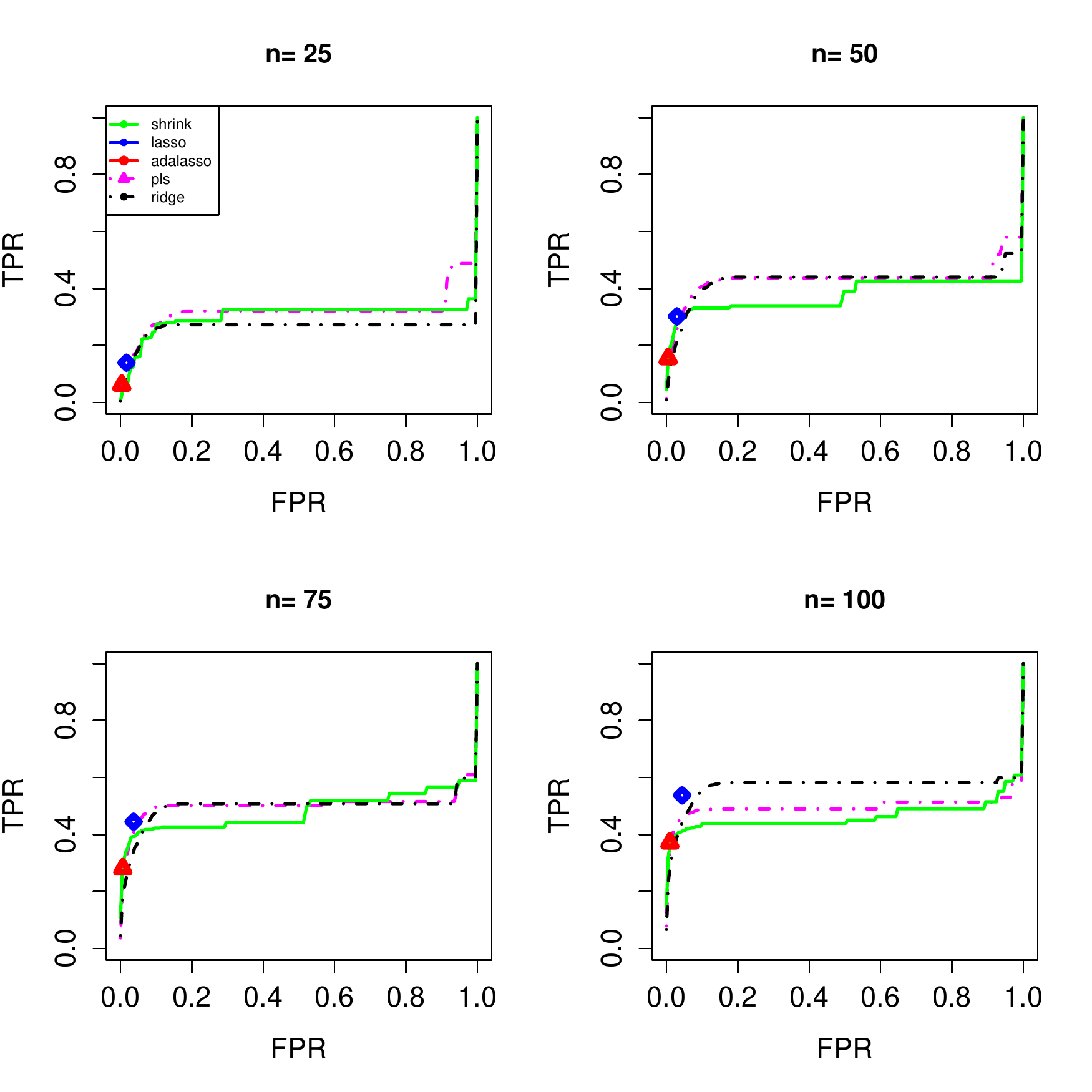}}}}\par}
\end{figure}

\newpage

\begin{figure}[htb]
{\par\centering\resizebox*{16cm}{16cm}{\rotatebox{0}{{\includegraphics{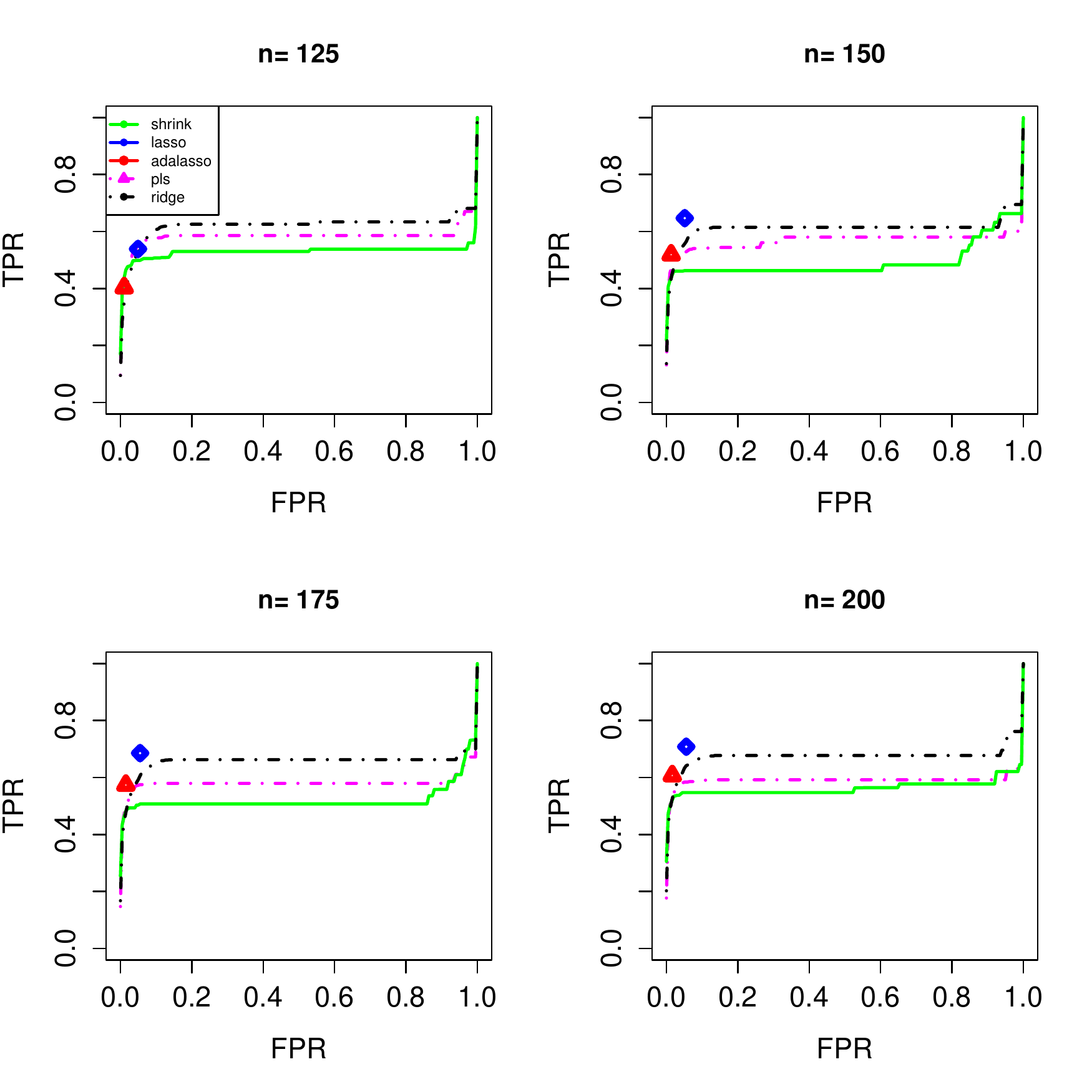}}}}\par}
\end{figure}

\newpage

\subsection*{Figure 2b - ROC curves for a density of $0.10$}
      ROC curves obtained by varying the fdr-threshold for PLS, Ridge Regression and Shrinkage. The sensitivity and specificity of Lasso and Adaptive Lasso are represented by a point.

\begin{figure}[htb]
{\par\centering\resizebox*{16cm}{16cm}{\rotatebox{0}{{\includegraphics{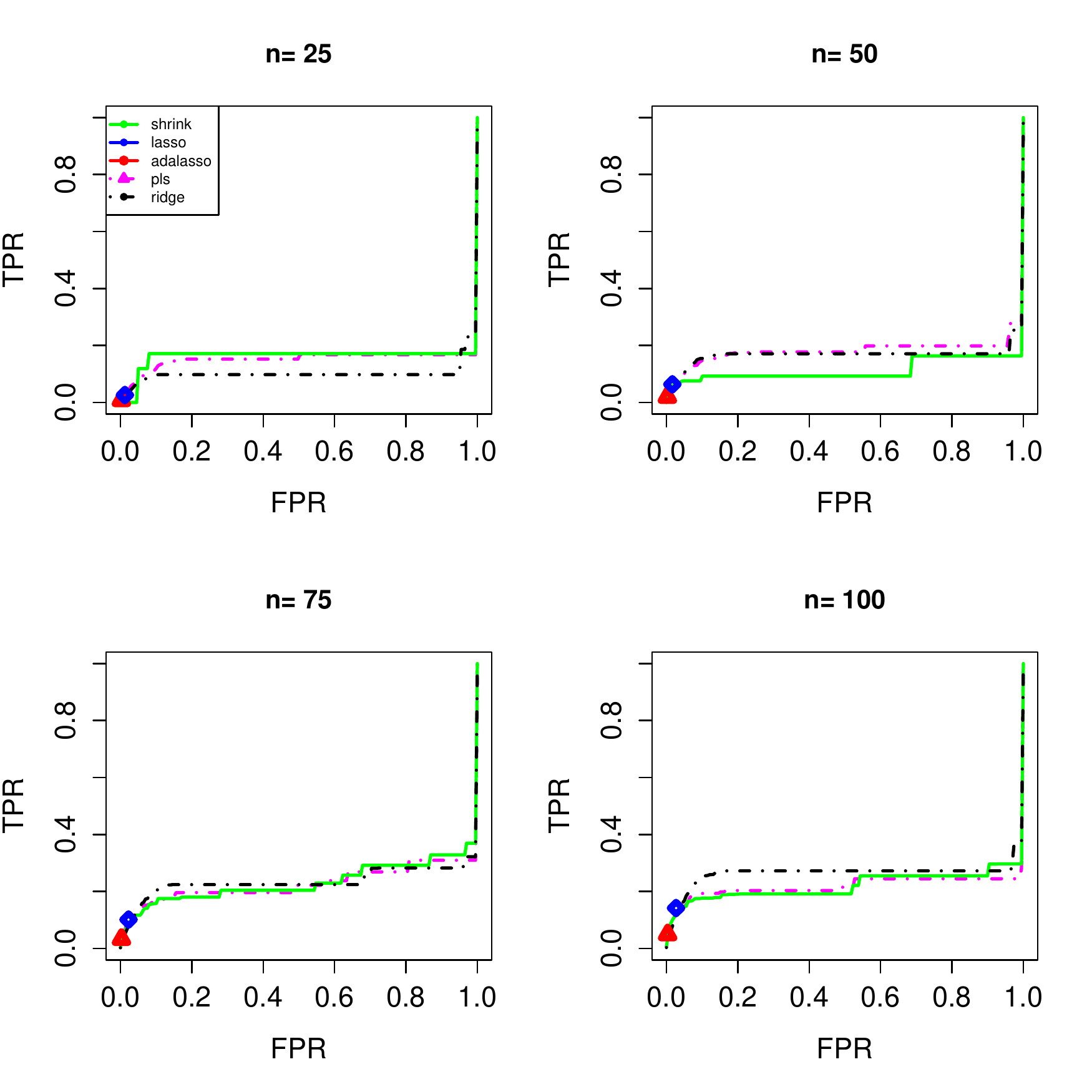}}}}\par}
\end{figure}

\newpage

\begin{figure}[htb]
{\par\centering\resizebox*{16cm}{16cm}{\rotatebox{0}{{\includegraphics{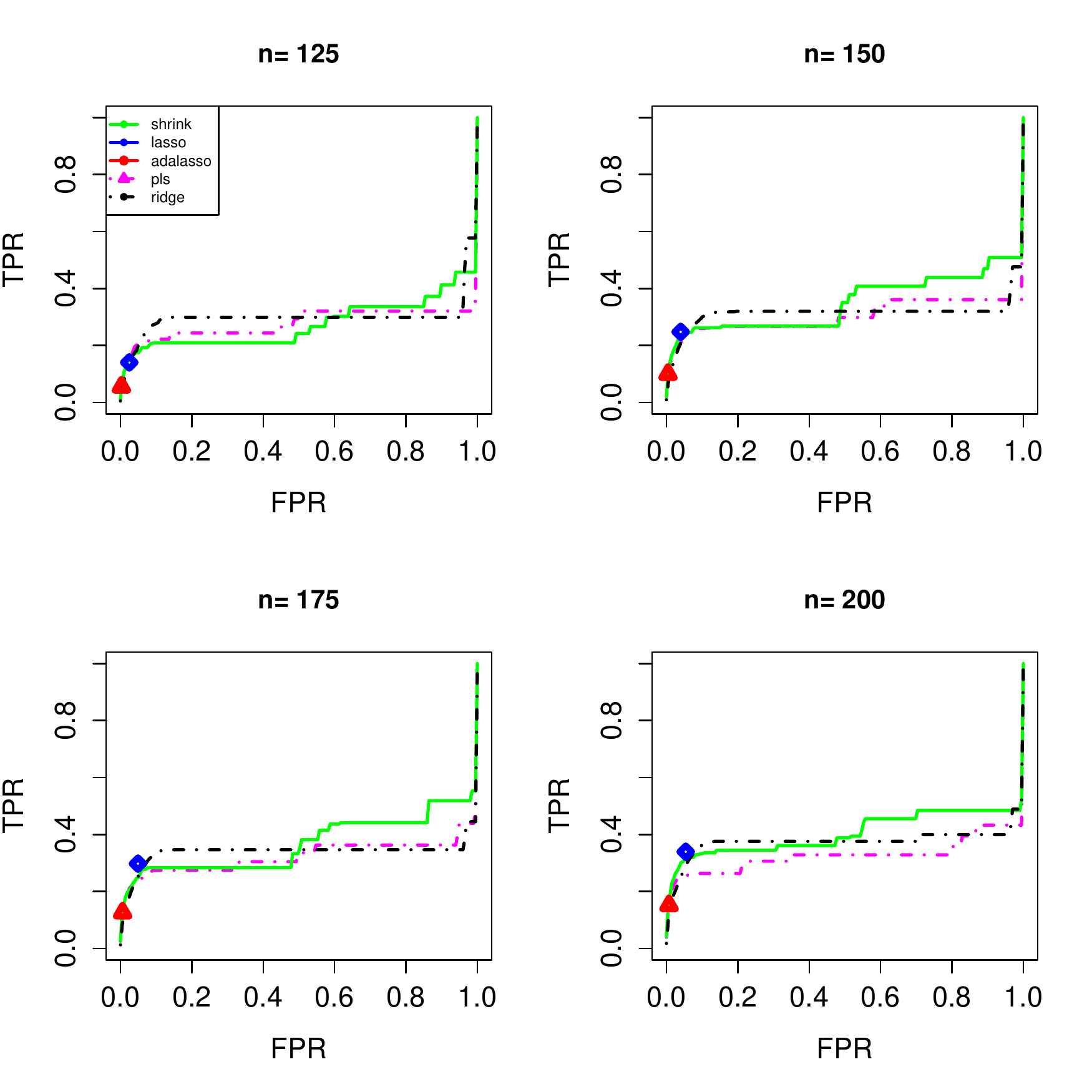}}}}\par}
\end{figure}

\newpage
\subsection*{Figure 2c - ROC curves for a density of $0.15$}
      ROC curves obtained by varying the fdr-threshold for PLS, Ridge Regression and Shrinkage. The sensitivity and specificity of Lasso and Adaptive Lasso are represented by a point.

\begin{figure}[htb]
{\par\centering\resizebox*{16cm}{16cm}{\rotatebox{0}{{\includegraphics{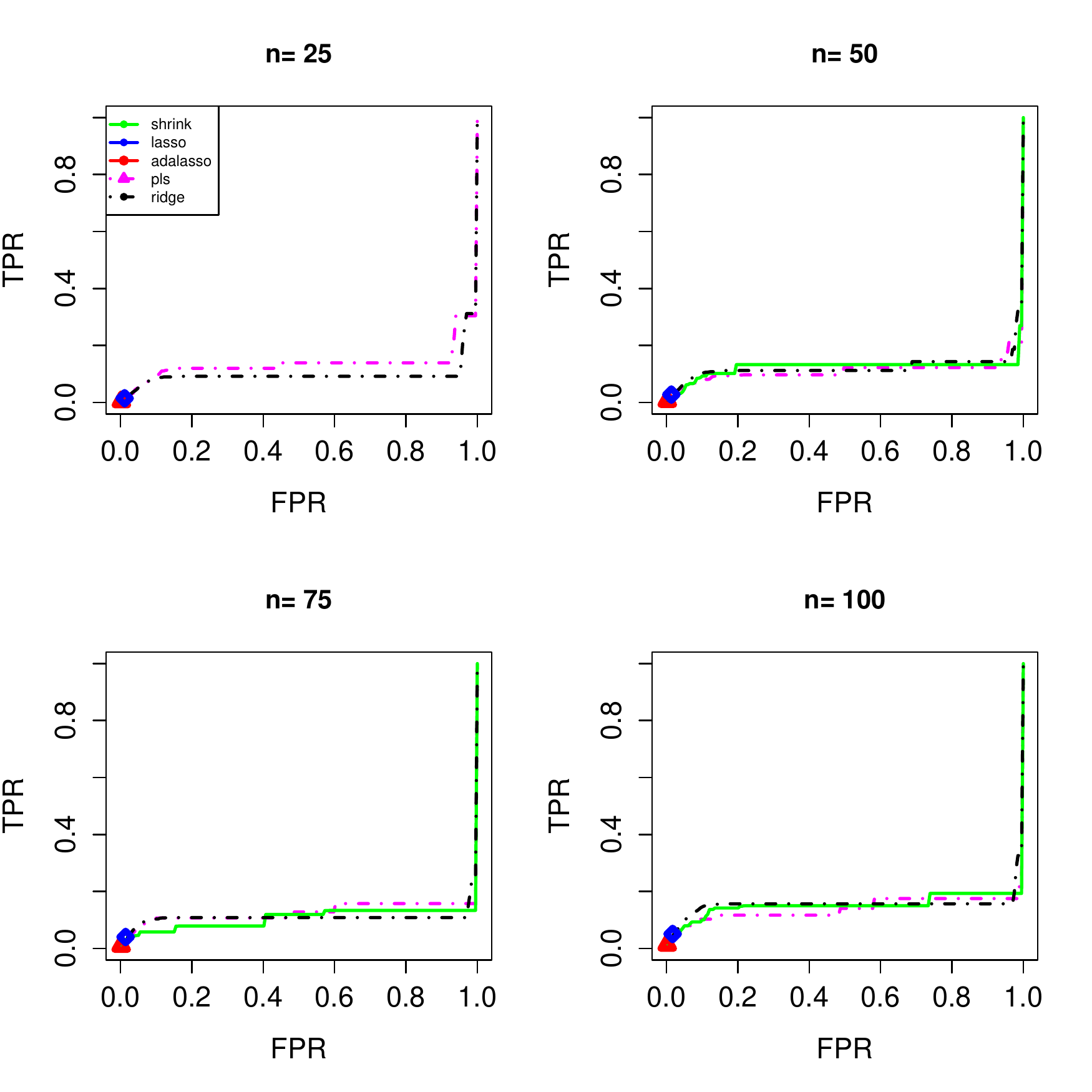}}}}\par}
\end{figure}

\newpage

\begin{figure}[htb]
{\par\centering\resizebox*{16cm}{16cm}{\rotatebox{0}{{\includegraphics{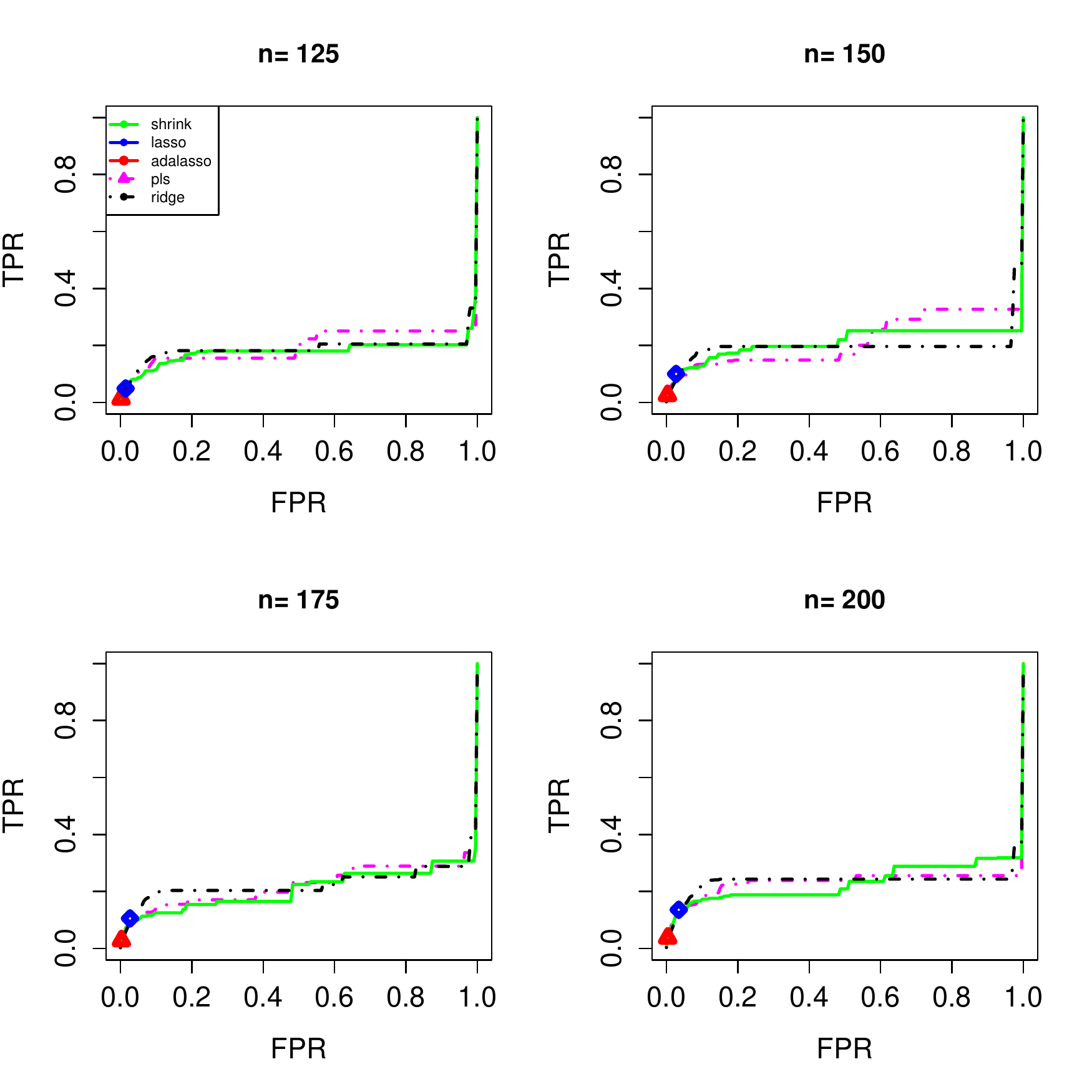}}}}\par}
\end{figure}

\newpage

\subsection*{Supplement 2d - ROC curves for a density of $0.20$}
      ROC curves obtained by varying the fdr-threshold for PLS, Ridge Regression and Shrinkage. The sensitivity and specificity of Lasso and Adaptive Lasso are represented by a point.

\begin{figure}[htb]
{\par\centering\resizebox*{16cm}{16cm}{\rotatebox{0}{{\includegraphics{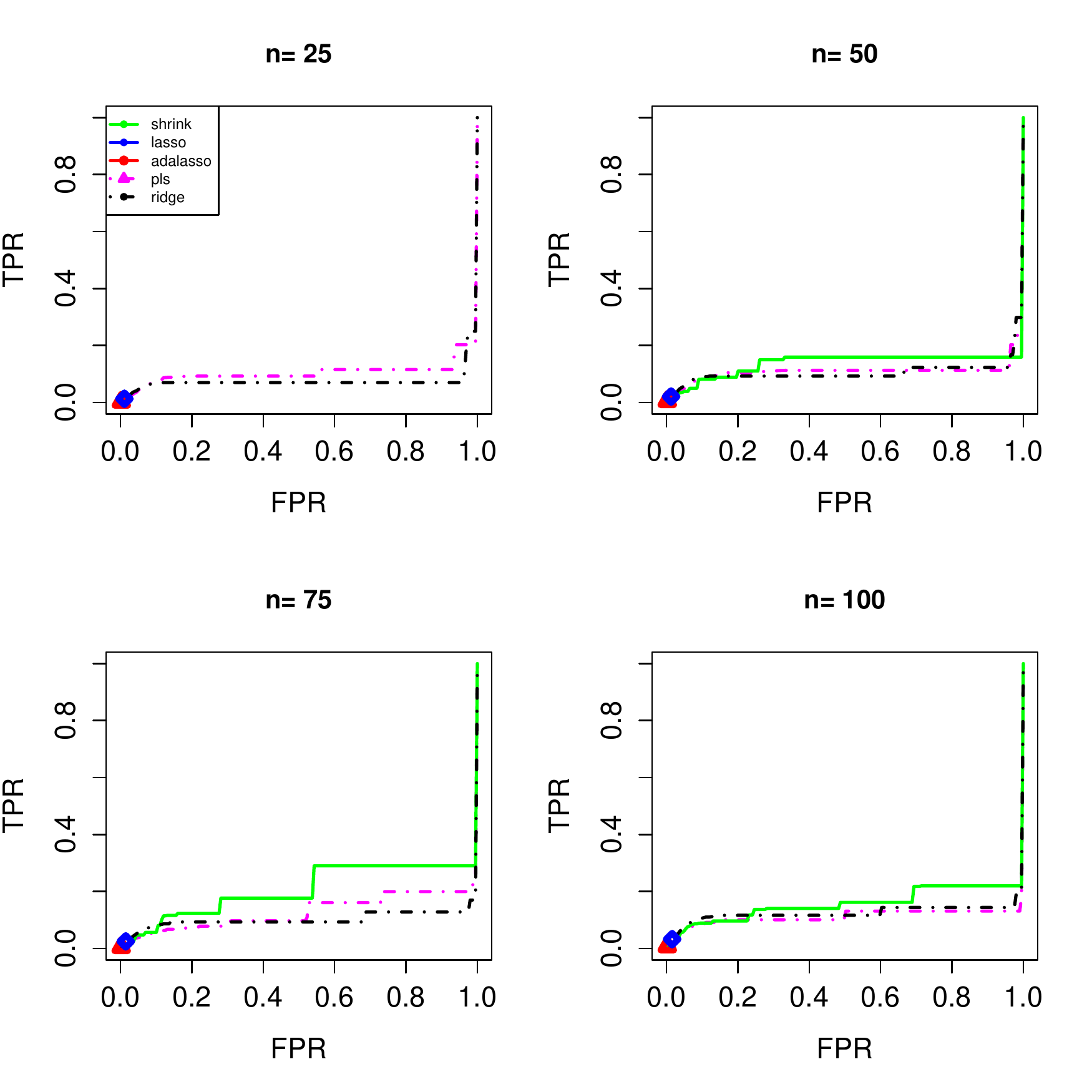}}}}\par}
\end{figure}

\newpage

\begin{figure}[htb]
{\par\centering\resizebox*{16cm}{16cm}{\rotatebox{0}{{\includegraphics{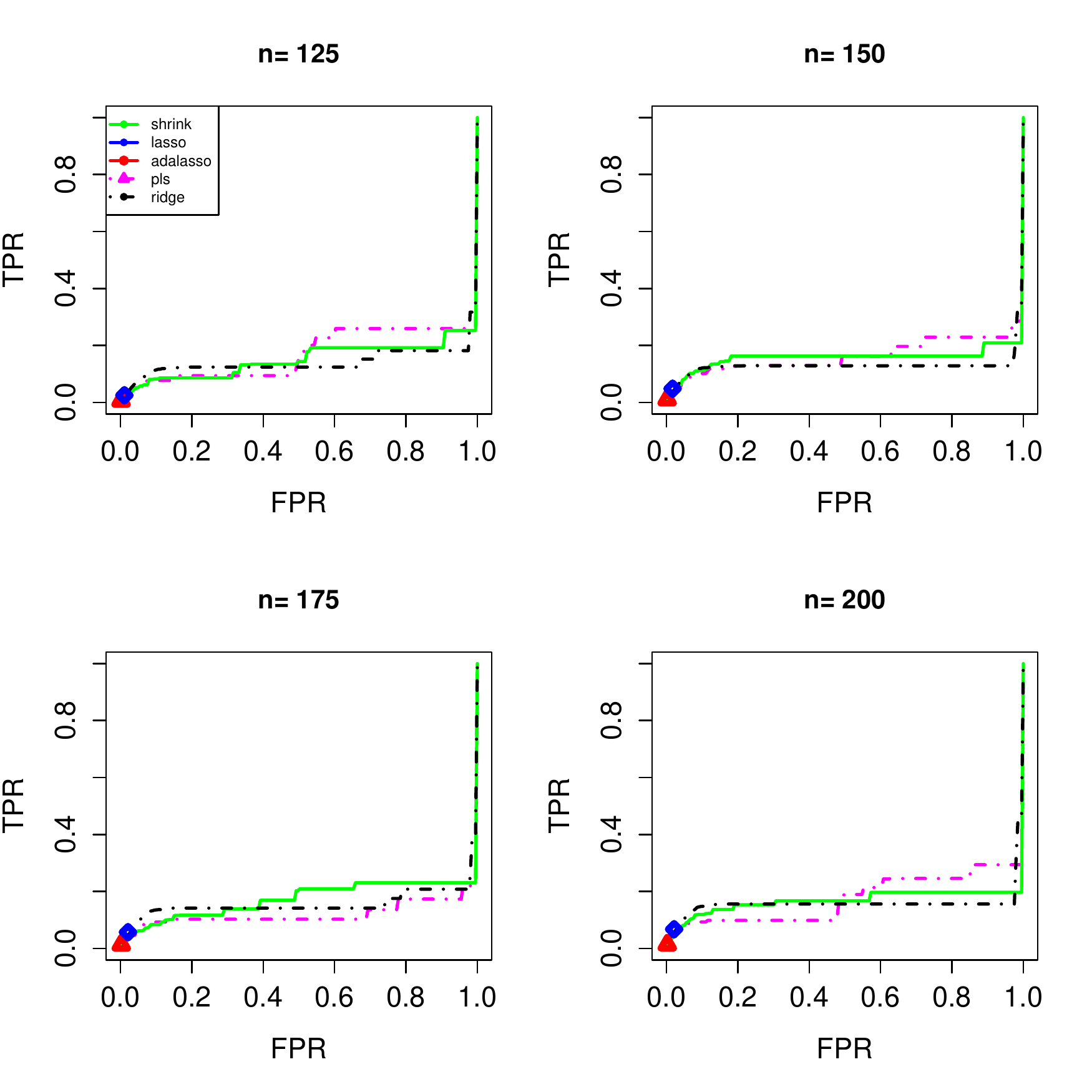}}}}\par}
\end{figure}

\newpage
\subsection*{Supplement 2e - ROC curves for a density of $0.25$}
      ROC curves obtained by varying the fdr-threshold for PLS, Ridge Regression and Shrinkage. The sensitivity and specificity of Lasso and Adaptive Lasso are represented by a point.

\begin{figure}[htb]
{\par\centering\resizebox*{16cm}{16cm}{\rotatebox{0}{{\includegraphics{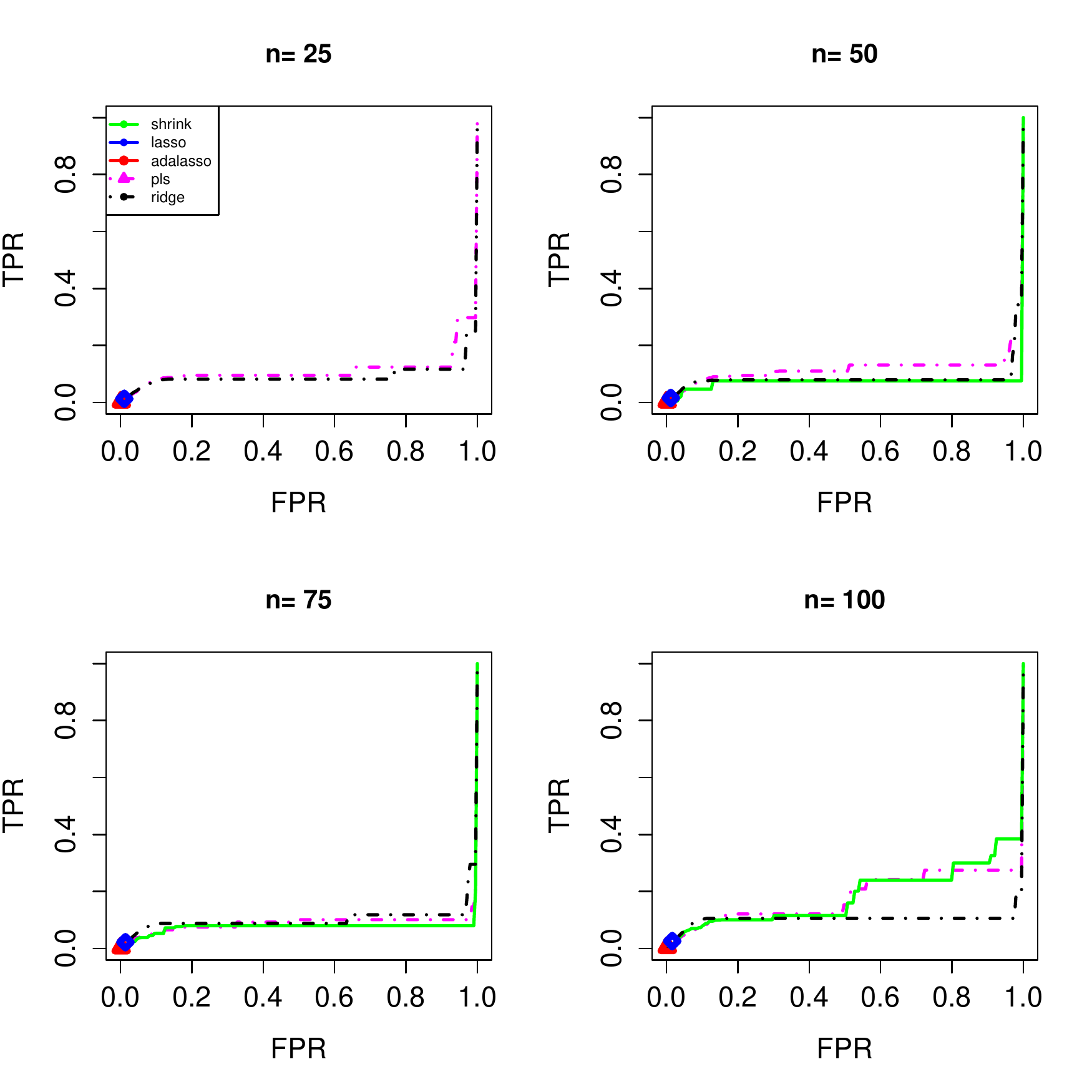}}}}\par}
\end{figure}

\newpage

\begin{figure}[htb]
{\par\centering\resizebox*{16cm}{16cm}{\rotatebox{0}{{\includegraphics{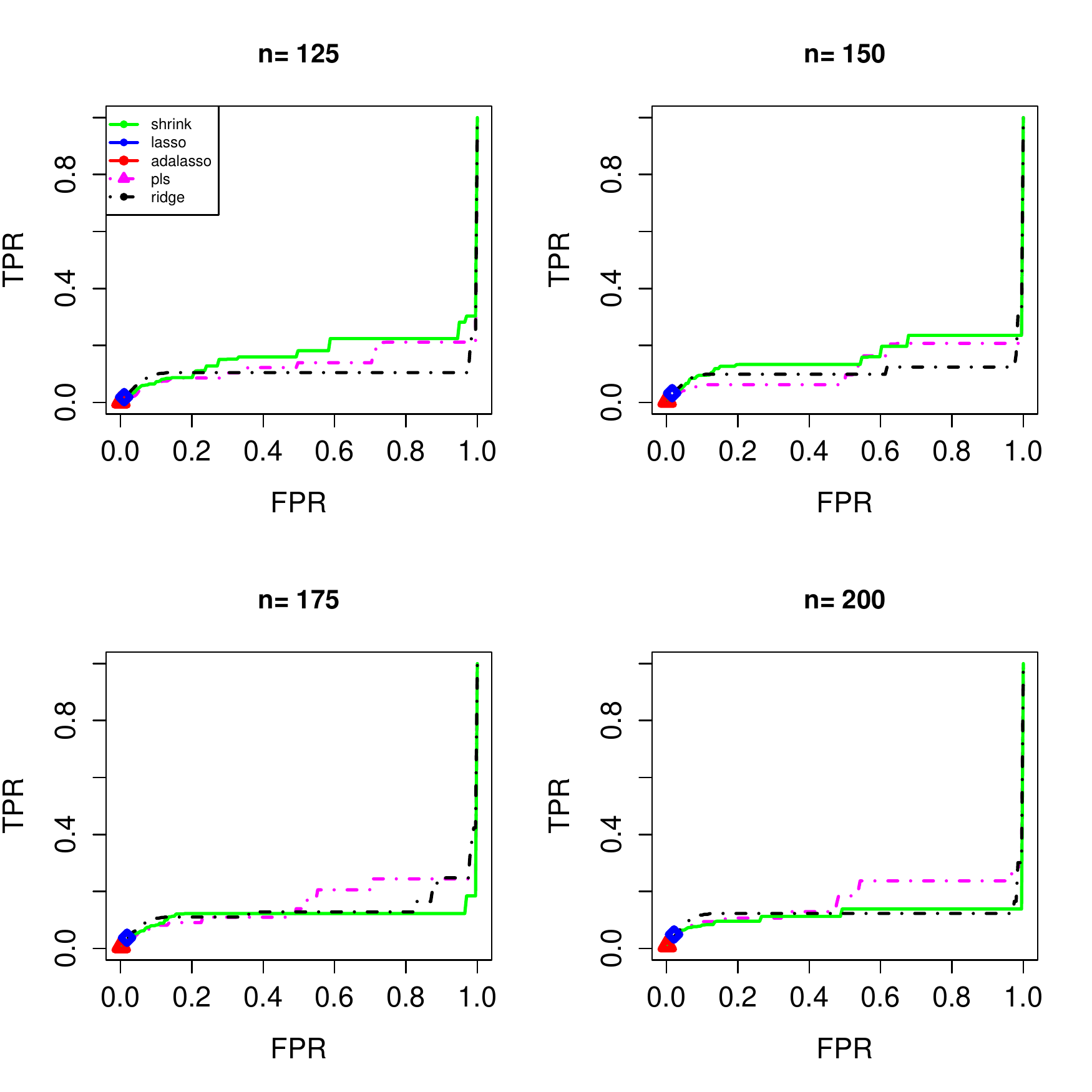}}}}\par}
\end{figure}
\newpage
\subsection*{Supplement 3 - Histogram of partial correlations for different degrees of density}

\begin{figure}[htb]
{\par\centering\resizebox*{16cm}{16cm}{\rotatebox{0}{{\includegraphics{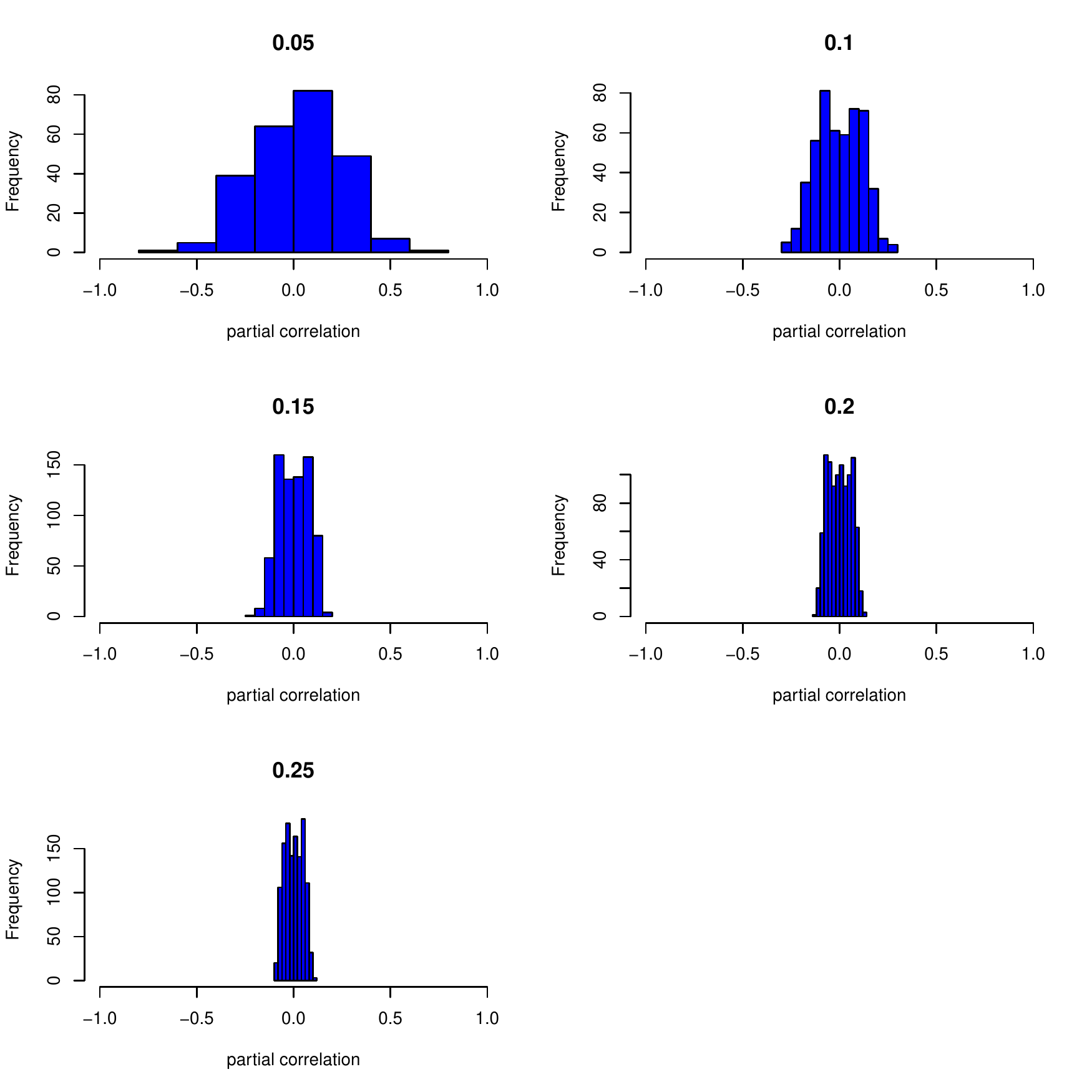}}}}\par}
\end{figure}
\newpage

\subsection*{Supplement 4 - Different network topologies}

\begin{figure}[htb]
{\par\centering\resizebox*{16cm}{8cm}{\rotatebox{0}{{\includegraphics{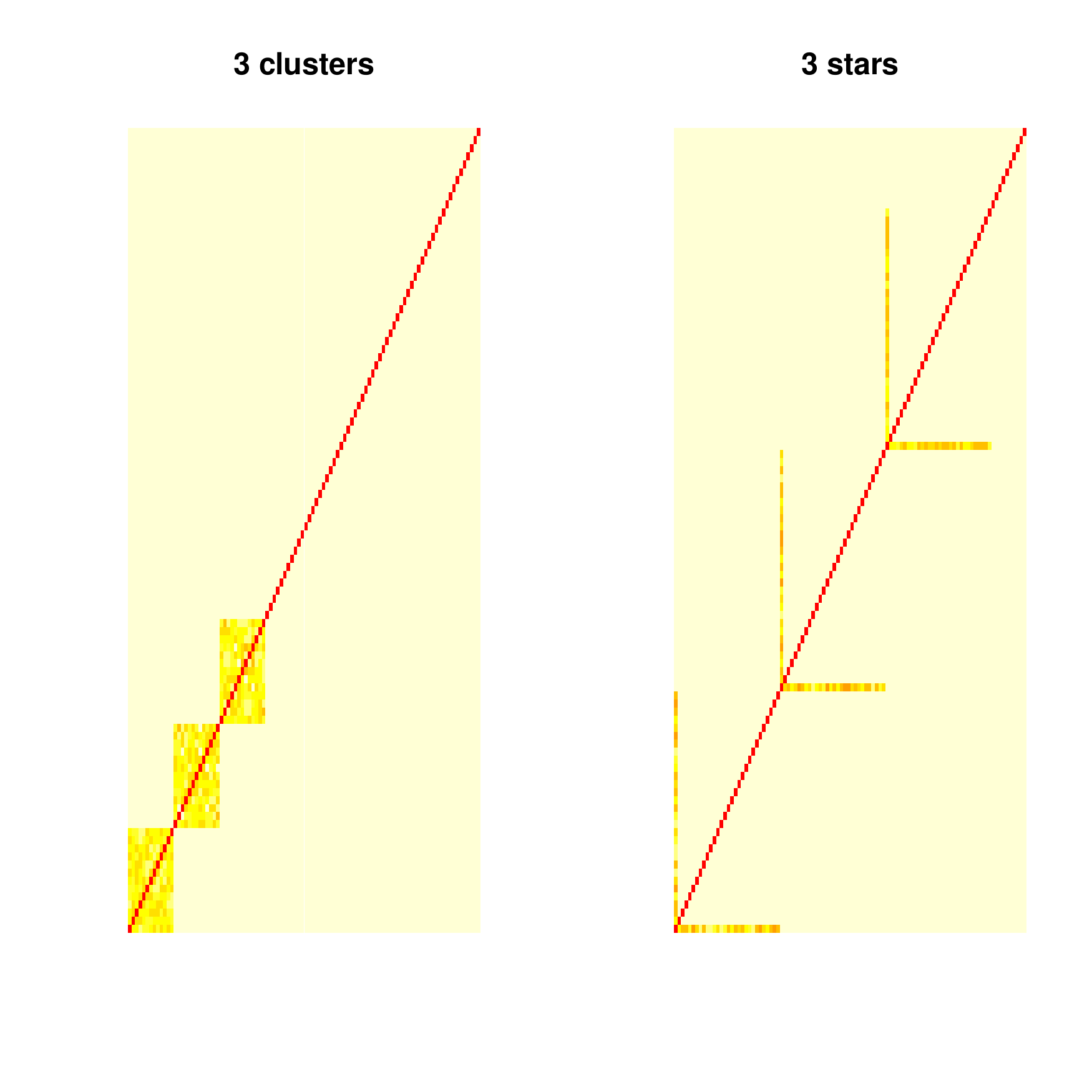}}}}\par}
\end{figure}

\end{document}